\documentclass{article}
\usepackage{enumitem}   

\usepackage{booktabs}

\RequirePackage[OT1]{fontenc}
\RequirePackage{amsthm,amsmath}
\RequirePackage[colorlinks,citecolor=blue,urlcolor=blue]{hyperref}

\usepackage{amsfonts}
\usepackage{mathrsfs}
\usepackage{graphics}
\usepackage{caption}
\usepackage{subcaption}
\usepackage{mathtools}
\usepackage{color}
\usepackage{floatrow}
\usepackage{array}
\usepackage{bbm}
\usepackage{booktabs}
\usepackage{float}
\usepackage[labelfont=bf]{caption}
\usepackage{varioref}
\usepackage{stackengine}

\usepackage{enumitem}
\usepackage[T1]{fontenc}
\usepackage{ucs}
\usepackage[utf8x]{inputenc}

\numberwithin{equation}{section}
\theoremstyle{plain}

\newtheorem{ex}{Example}[section]

\newcommand{\HH}{\mathcal H}

\newcommand{\R}{\ensuremath{\mathbb{R}}}

\renewcommand{\P}{\mathbb P}

\newcommand{\E}{\ensuremath{\mathbb{E}}}

\renewcommand{\H }{\"}

\renewcommand{\H}{\boldsymbol{h}}
\renewcommand{\S}{\boldsymbol{s}}
\renewcommand{\HH}{\ensuremath{\lVert \boldsymbol{h} \rVert} }

 \usepackage{csquotes}

\begin{document}

\title{Semiparametric estimation for space-time max-stable processes: $F$-madogram-based estimation approach}

\author{Abdul-Fattah Abu-Awwad, abuawwad@math.univ-lyon1.fr \& \\
V\'{e}ronique Maume-Deschamps, veronique.maume@univ-lyon1.fr \& \\
Pierre Ribereau, pierre.ribereau@univ-lyon1.fr\\
Universit\'{e} de Lyon, Universit\'{e} Claude Bernard Lyon 1,\\
Institut Camille Jordan ICJ UMR 5208 CNRS, France.}

\date{}

\maketitle


\begin{abstract}
Max-stable processes have been expanded to quantify extremal dependence in spatio-temporal data. Due to the interaction between space and time, spatio-temporal data are often complex to analyze. So, characterizing these dependencies is one of the crucial challenges in this field of statistics. This paper suggests a semiparametric inference methodology based on the spatio-temporal $F$-madogram for estimating the parameters of a space-time max-stable process using gridded data. The performance of the method is investigated through various simulation studies. Finally, we apply our inferential procedure to quantify the extremal behavior of radar rainfall data in a region in the State of Florida. 
\end{abstract}

%
%

\section{Introduction}
\begin{figure} [!htp]
	\centering
	\includegraphics[scale=0.45]{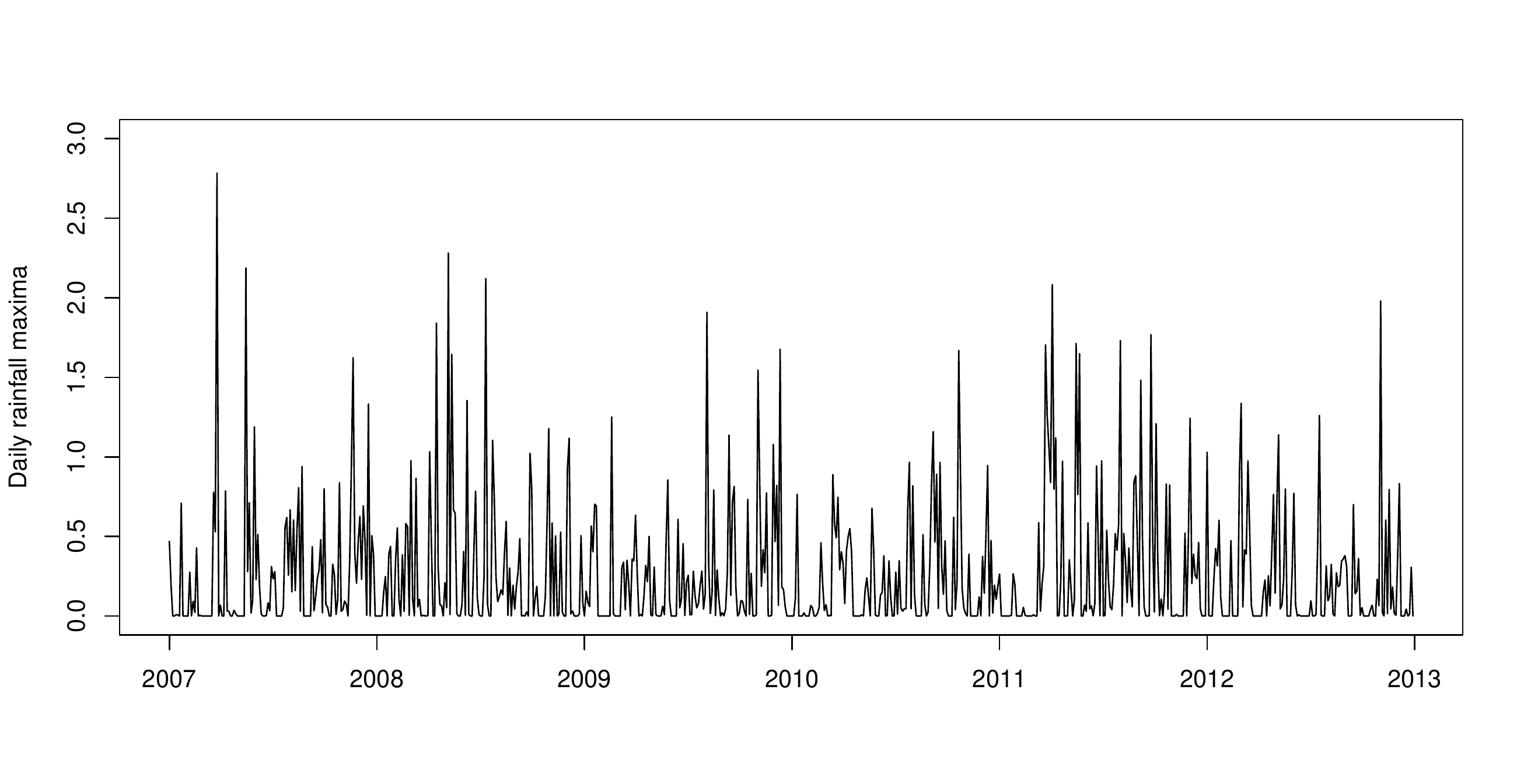}
	
	\caption{Daily rainfall maxima in inches taken over hourly measurements from 2007-2012 for a fixed location in Florida, USA.}
	\label{Dailyint}	
\end{figure}

Typically, extremes of environmental and climate processes like extreme wind speeds or heavy precipitation are modelled using extreme value theory. Max-stable processes are ideally suited for the statistical modeling of spatial extremes as they form the natural extension of multivariate extreme value distributions to infinite dimensions. Various families of max-stable models and estimation procedures have been proposed for extremal data. For a detailed overview of max-stable processes, we refer the reader to \cite{de2007extreme}. For statistical inference, it is then often assumed that the observations at spatial locations are independent in time, see, e.g., \cite{padoan2010likelihood, davison2012statistical, davison2013geostatistics}. However, many extreme environmental processes observations exhibit a spatial dependence structure, meaning that neighboring locations within some distance show similar patterns, as well as a temporal dependence, which can be seen from high values for two consecutive time moments (e.g., within hours). As an illustration, Figure~\ref{Dailyint} depicts the daily rainfall maxima for the wet seasons (June-September) from the years 2007-2012 at one fixed grid location in Florida. We observe that it is likely that a high value is followed by a value of a similar magnitude. So, the temporal dependence may be present. Accordingly, the temporal dependence structure should be considered in an appropriate way. More details on the rainfall data in Florida are given in Section \ref{sec:real}.

Currently space-time models are still taking up little space in the literature. Only a few papers have introduced space-time max-stable models. For instance, \cite{davis2013max} extended the construction of spatial Brown-Resnick (BR) model \cite{brown1977extreme, kabluchko2009stationary} and Smith’s storm profile model \cite{smith1990max} to the space-time domain, whereas \cite{buhl2016anisotropic} extended the space-time BR model \cite{davis2013max} to an anisotropic setting. Additionally, \cite{huser2014space} introduced an extension of spatial Schather model \cite{schlather2002models}, which comprises a truncated Gaussian random process, so that storm shapes are stochastic, and includes a compact random set, that allows the process to be mixing in space as well as to exhibit a spatial diffusions, see also \cite{davison2012geostatistics}. A common feature for these models is that the major emphasis is in modeling asymptotic dependence treating the time just as additional dimension of the space. So, these models do not allow any interaction between the spatial components and temporal component in the underlying dependence function. However it seems reasonable to suppose that the
spatial and temporal components behave asymptotically in a different way. Therefore, a new class of space-time max-stable models have been proposed by \cite{embrechts2016space} in which the influence 
of time and space are partly decoupled, where the time infuences space through a bijective operator on space.\\
The inference on max-stable processes in both spatial and spatiotemporal contexts is an open field that is still in development. Many techniques have been proposed for parameter estimation in spatial extreme models. Each technique has its pros and cons. As with spatial max-stable processes, the pairwise likelihood estimation has been found useful to estimate the parameters of space-time max-stable processes due to its theoretical properties, see, e.g. \cite{davis2013statistical, huser2014space, embrechts2016space}. Recently, various semiparametric estimation approaches have been proposed to fit such processes. For instance \cite{buhl2016semiparametric} introduced a new semiparametric estimation procedure based on a closed form expression of the so-called extremogram \cite{davis2009extremogram} to estimate the parameters of space-time max-stable BR process. The extremogram has been estimated nonparametrically by its empirical version,
where space and time are separated. A constrained weighted linear regression is then applied in order to
produce parameter estimates. While in \cite{manaf} a semiparametric estimation procedure has been developed for spatial max-mixture processes \cite{doi:10.1093/biomet/asr080} based on the $F$-madogram \cite{cooley2006variograms}. A non-linear least squares (NLS) is then applied to minimize the squared difference between the empirical $F$-madogram and its model-based counterpart. A major advantage of the semiparametric methods is the substantial reduction of computation time compared to the pairwise likelihood estimation. Hence, these methods can be applied as an alternative or a prerequisite to the widely-adopted pairwise likelihood inference, which suffers from some defects; first, it can be onerous, since the computation and subsequent optimization of the objective function is time-consuming. Second, the choice of good initial values for optimization of the composite likelihood is essential.

An implicit difficulty in any extreme value analysis is the limited amount of data for model estimation, see, e.g. \cite{coles2001introduction}. Hence, inference based on the extremogram is difficult because few observations are available as the threshold increases. Consequently, the semiparametric estimates obtained by \cite{buhl2016semiparametric} showed a larger bias than the pairwise likelihood estimates and are sensible to the choice of the threshold used for the extremogram. Accordingly, the surrogates of existing estimation techniques should be welcomed.

In the present paper, we are interested in statistical inference for space-time max-stable processes. Motivated by deficiencies in existing inference approaches, we propose two novel and flexible semiparametric estimation schemes to fit space-time max-stable processes:
\begin{enumerate}[label=(\roman*)]
	\item \textbf{Scheme 1:} we estimate spatial and temporal parameters separately. Based on NLS, we minimize the squared difference between the empirical estimates of spatial/temporal $F$-madograms and their model-based counterparts. Our inferential methodology is close to the one that has been proposed by \cite{buhl2016semiparametric} as an alternative or a preliminary analysis to the pairwise likelihood approach in \cite{davis2013statistical}, where only isotropic space-time max-stable BR process has been fitted via the two approaches. 
	
	\item \textbf{Scheme 2:} we generalize the NLS to estimate spatial and temporal parameters simultaneously. 
\end{enumerate}

The remainder of the paper is organized as follows. Section \ref{sec:models} defines the space-time max-stable models. The two semiparametric estimation schemes are described in Section \ref{sec:inference}. Section \ref{sec:simulation} illustrates the performance of our method through various simulation studies, where also a comparison with the semiparametric estimation \cite{buhl2016semiparametric} is performed. In Section \ref{sec:real}, we apply our method to radar rainfall data in a region in Florida by using spatial and temporal block maxima design. The concluding remarks in Section \ref{sec:conc} address some remaining issues and perspectives. 

\section{Space-time max-stable models } \label{sec:models}
Throughout the paper, $X:=\left\{ X(\S,t): (\S,t) \in \mathcal{S}\times \mathcal{T} \right\}$, $\mathcal{S}\times \mathcal{T} \subset \R^d \times \R^+$ (generally, $d=2$) is a spatiotemporal process, where the
space $\mathcal{S}\times \mathcal{T}$ is the spatiotemporal domain. The points $\S$ denote the spatial coordinates and are called “sites” or “locations” or “stations” and the points $t$ denote the temporal coordinates and are called “times” or “moments”. The space index $\S$ and time index $t$ will respectively belong to the sets $\mathcal{S}$ and $\mathcal{T}$. In addition, we will denote by $\H=\S_1-\S_2 \in \R^2$ (respectively $l=t_1-t_2 \in \R$) the spatial (respectively temporal) lag.

\subsection{Space-time max-stable models without spectral separability}
According to \cite{de1984spectral}, the simple space-time max-stable  process $X$, where simple means that the margins are standard Fr\'{e}chet, i.e., $F(x):=\mathbb{P} (X(\S,t)\leq x)= \exp \{-x^{-1}\}$, $x>0$, has the following spectral representation 
	\begin{equation}
X(\boldsymbol{s},t)\stackrel{\mathcal{D}}{=}  \bigvee_{i=1}^{\infty} {\xi_{i} U_i {(\boldsymbol{s},t)}}, \   (\boldsymbol{s},t) \in \mathcal {S}\times \mathcal{T}.
\label{spectral rep}
\end{equation}

where $\bigvee$ denotes the max-operator, $\{\xi_{i}\}_{i \geq 1}$ are independent and
identically distributed (i.i.d.) points of a Poisson process on $(0,\infty)$ with intensity $\xi^{-2} d\xi$ and $\{U_{i} (t,\S)\}_{i \geq 1}$ is a sequence of independent replications of some space-time process $\{U(\S,t), (\S,t) \in \mathcal {S}\times \mathcal{T}\}$ with $\E\{U(\S,t)\} < \infty$ for each $(t,\boldsymbol{s}) \in \mathcal {S}\times \mathcal{T}$, and $U(\S,t) \geq 0$, which are also independent of $\xi_{i}$.

For $D \in \mathbb{N}\setminus \{0\}$, $\boldsymbol{s}_{1},\ldots,\boldsymbol{s}_{D} \in \mathcal{S}$, $t_{1},\ldots,t_{D} \in \mathcal{T}$ and $x_{1},\ldots,x_{D}>0$, the finite $D$-dimensional distributions of the space-time max-stable process $X$ are given by
\begin{align} 
\mathbb{P}(X(\S_1,t_1)\leq x_{1},\ldots,X(\S_D,t_D)\leq x_{D})=& \mathbb{P} \left\{ \xi_i \bigvee_{j=1}^{D} \frac{U_i {(\S_{j},t_j)}}{x_j} \leq 1,  \forall i=1,2,\ldots  \right\} \\  \nonumber
&=\exp \left\{ -\mathbb{E} \left(\bigvee_{j=1}^{D} \frac{U {(\boldsymbol{s}_{j},t_j)}}{x_j} \right)  \right\}. 
\end{align}
Hence, all finite-dimensional distributions are multivariate extreme value distributions with unit Fr\'{e}chet margins. In particular, for $x_1,x_2 > 0$, the bivariate cumulative distribution function (c.d.f.) $F_{\S_1,t_1,\S_2,t_2}$ of the space-time max-stable process $X(\S,t)$ in (\ref{spectral rep}) can be expressed in terms of the underlying bivariate spatio-temporal exponent function $V_{\S_1,t_1,\S_2,t_2}$ as
\begin{align} \label{biv exponent measure}
- \log F_{\S_1,t_1;\S_2,t_2}(x_1,x_2)=&- \log \mathbb{P}\left[X(t_1,\boldsymbol{s}_1) \leq x_{1}, X( t_2,\boldsymbol{s}_2) \leq x_{2} \right]\\  \nonumber & =: V_{\S_1,t_1;\S_2,t_2}\left(x_1,x_2\right).
\end{align}
Below, we will consider stationary space-time processes, so that $V_{\S_1,t_1;\S_2,t_2}$ depends only on $\H=\S_1-\S_2$ and $l=t_1-t_2$. We will write $F_{\H,l}$ for $F_{\S_1,t_1;\S_2,t_2}$ and $V_{\H,l}$ for $V_{\S_1,t_1;\S_2,t_2}$.

\subsubsection{{Spatio-temporal extremal dependence summary measures}}
In order to measure the spatio-temporal extremal dependence, we provide in the next Definition, extensions to the spatio-temporal setting of some quantities that have been introduced in the spatial context. For a stationary spatio-temporal max-stable process $X$ with univariate margin c.d.f. $F$, we have
\begin{enumerate}[label=(\roman*)]
	\item \textbf{(Spatio-temporal extremal dependence function, originally due to \cite{schlather2003dependence}) }
	\begin{equation}	\label{Def::spatiotemporal-theta} 
	\theta(\H,l)= - x\log \P \left( X(\S,t) \leq x, X(\S+\H,t+l) \leq x \right) \in [1,2], \ x>0.
	\end{equation}
	\item \textbf{(Spatio-temporal upper tail dependence function, originally due to \cite{coles1999dependence})}
	\begin{equation} \label{Def::spatiotemporal-chi}
	\chi_u{(\H,l)}= 2- \frac{2 \log \mathbb{P}\{F(X(\S,t))<u , F(X(\S+\H,t+l))<u\}}{\log \mathbb{P}\{F(X(\S+\H,t+l))<u\}}
	\end{equation}
	and $\chi{(\H,l)}=\lim_{u\rightarrow 1^-}\chi_u{(\H,l)}$, $u \in [0,1]$. Similarly to spatial setting, we have the simple link: $\chi{(\H,l)}=2-\theta(\H,l)$.

	\item \textbf{(Spatio-temporal $F$-madogram, originally due to \cite{cooley2006variograms})}	
	\begin{equation}
	\nu_{F}(\H,l)= \frac{1}{2} \mathbb{E} \left[|F(X( \S,t )) - F(X( \S+\H,t+l ))|\right] \in [0,1/6].
	\label{Def::spatiotemporal-mado} 
	\end{equation}
	Furthermore, the $F^\lambda$-madogram (originally due to \cite{bel2008assessing}) and $\lambda$-madogram (originally due to \cite{naveau2009modelling}) are defined analogously.

	\item \textbf{(Spatio-temporal extremogram dependence function, originally due to {\cite{davis2009extremogram}})}
	\begin{equation}\label{Space-time::Extremo}
	\rho_{\mathscr{A}_1, \mathscr{A}_2}(\H,l)= \lim_{x \rightarrow \infty} \frac{ \mathbb{P}\left\{x^{-1} X(\S,t) \in \mathscr{A}_1 , x^{-1}X(\S+\H,t+l) \in \mathscr{A}_2\right\}}{ \mathbb{P}\left\{ x^{-1} X(\S,t) \in \mathscr{A}_1\right\}}. 
	\end{equation} 
	Clearly, setting the Borel sets $\mathscr{A}_1=\mathscr{A}_2=(1,\infty)$ yields $\rho_{(1,\infty), (1,\infty)}(\H,l) =\chi(\H,l)$. The two cases $\chi(\H,l)=0$ and $\chi(\H,l)=1$ correspond to the boundary cases of asymptotic independence and complete dependence.
\end{enumerate}
Both dependence functions $\theta(\H,l)$ and $\chi(\H,l)$ provide simple measures of extremal dependence within the class of asymptotic dependence distributions.

\begin{ex} \textbf{(Stationary BR spatio-temporal process without spectral separability)}
	\cite{davis2013max} introduced the spatial BR model \cite{brown1977extreme, kabluchko2009stationary} in space and time. A strictly stationary spatio-temporal BR process $X$ has the following spectral representation
\begin{equation}
X(\S,t)= \bigvee_{i=1}^{\infty} \xi_i \exp \left\{\varepsilon_i(\S,t) - \gamma(\S,t) \right\}, \ (\S,t) \in \mathcal {S}\times \mathcal{T},
\label{BR spectral fun}
\end{equation}
where $ \left\{\xi_{i}\right\}_{i \geq 1}$ are points of a Poisson process on $(0,\infty)$ with intensity $\xi^{-2} d\xi$, the processes $\left\{\varepsilon_i(\S,t): {(\S,t)\in (\mathcal{S}\times \mathcal{T})}  \right\}$ are independent replications of a Gaussian process $\{\varepsilon(t,\S)\}$ with stationary increments, $\varepsilon(\boldsymbol{0},0)=0$, $\mathbb{E}[\varepsilon(\S,t)]=0$ and covariance function $$ \mathbb{C}ov (\varepsilon(\S_1,t_1),\varepsilon(\S_2,t_2) ) = \gamma(\S_1 ,t_1) + \gamma(\S_2,t_2) - \gamma( \S_1-\S_2 , t_1-t_2),$$
for all $( \S_1,t_1),(\S_2,t_2) \in\mathcal {S} \times \mathcal{T}$.
The dependence function $\gamma$ which is termed the spatio-temporal semivariogram of the process $\{\varepsilon(\S,t)\}$, is non-negative and conditionally negative definite, that is,
for any $k \in \mathbb{N}$, $(\S_1,t_1),\ldots,(\S_k,t_k) \in \mathcal{S}\times\mathcal{T} $ and $a_1, \ldots,a_k \in \mathbb{R}$, 
$$ \sum_{i=1} ^{k} \sum_{j=1} ^{k} a_i a_j \gamma \left( \S_i -\S_j ,t_i-t_j \right) \leq 0,\ \sum_{i=1}^{k}a_i =0.$$

The process $X(\S,t)$ in (\ref{BR spectral fun}) is fully characterized by the dependence function $\gamma$. In geostatistics, the function $\gamma$ 
is given by 
$$ \gamma\left( \S_1-\S_2,t_1-t_2 \right)= \frac{1}{2} \mathbb{V}ar \left(\varepsilon(\S_1,t_1)-\varepsilon(\S_2,t_2) \right).$$ 
Let $\Phi$ denote the standard normal distribution function. For $x_1,x_2>0$, the bivariate c.d.f. ($F_{\H,l}$) of $\left(X(\S_1,t_1),X(\S_2,t_2)\right)$ in the stationary case is given by
\begin{align} \label{Biv BR}
- \log F_{\H,l}(x_1,x_2) = \frac{1}{x_{1}}\Phi\left(\sqrt{\frac{\gamma({\H}, l)}{2}}+\frac{\log\left(\frac{x_{2}}{x_{1}}\right)}{\sqrt{2\gamma({\H} ,l )}}\right )\\+ \nonumber
\frac{1}{x_{2}}\Phi\left( 
\sqrt{\frac{\gamma( {\H},l)}{2}}+\frac{ \log\left(\frac{x_{1}}{x_{2}}\right)}{\sqrt{2\gamma( {\H},l)}}\right).
\end{align}
Recall that if $\gamma$ is assumed to depend only on the norm of $\S_1-\S_2$, the associated process is spatially isotropic. The pairwise spatio-temporal extremal dependence function for this model is  $\theta({\H},l)=2\Phi\left\{\sqrt{\gamma({\H},l)/2}\right\}$. This model has been used in \cite{buhl2016semiparametric} to quantify the extremal behavior of radar rainfall data in a region of Florida, where a new semiparametric procedure based on the extremogram is applied to estimate the model parameters.
\end{ex}

\subsection{Space-time max-stable models with spectral separability}

The fundamental advantages of the spectral representation in (\ref{spectral rep}) are (i) the  construction of spatio-temporal processes from widely studied max-stable processes (ii) the huge literature available on spatio-temporal correlation functions for Gaussian processes, allows for considerable diversity of spatio-temporal behavior. However, an important modeling issue is that they do not allow any interaction between the spatial and the temporal components in the underlying dependence function. Thus, the time has no specific role but is equivalent to an additional spatial dimension; the spatial and temporal distributions belong to a similar family of models. Hence, alternatively, a new class of space-time max-stable models with spectral separability has been suggested in \cite{embrechts2016space}. More precisely,  
\begin{equation}
X(\S,t)=  \bigvee_{i=1}^{\infty} \xi_{i}U_ {t}(Q_i) U_ {\mathcal{R}(t,Q_i){\S}}(W_i), 
\label{spectral rep1}
\end{equation}
 where $\{\xi_{i},Q_i,W_i\}_{i \geq 1}$ are the points of a Poisson process on $(0,\infty) \times E_{1} \times E_{2} $, and with intensity  $\xi^{-2} d\xi \times \mu_{1} (dq)  \times \mu_{2} (dw)$ for some Polish measure spaces $(E_{1},\mathcal{E}_{1},\mu_{1})$ and $(E_{2},\mathcal{E}_{2},\mu_{2})$. The spectral function $U_ {t}:E_{1} \rightarrow (0,\infty)$ is measurable such that $\int_{E_{1}} U_ {t}(q)\mu_{1} (dq) =1$ for each $t \in \mathcal {T}$ and contributes to the temporal dynamic of the process, whereas the spectral function $U_ {\S}:E_{2} \rightarrow (0,\infty)$ is measurable such that $\int_{E_{2}} U_ {\S}(w)\mu_{2} (dw) =1$ for each $\boldsymbol{s} \in \mathcal {S}$ and drives the shape of the main spatial patterns. The operators $\mathcal{R}(t,q)$ are bijective from $\mathcal{S}$ to $\mathcal{S}$ for each $(t,q) \in \mathcal {T}\times E_{1 }$ and describes how the spatial patterns move in space.

The construction (\ref{spectral rep1}) allows one to deal with the temporal and spatial aspects separately. So, the estimation procedure can be simplified by estimating in a first step the spatial parameters independently from the temporal ones. Several examples of subclasses of the general class of space-time process $X$ (\ref{spectral rep1}) were introduced by \cite{embrechts2016space}, where the operator is either a translation or a rotation. The authors in that paper focused mainly on a special case of models where the function corresponding to the time in the spectral representation is the exponential density (continuous-time case) or the probability values of a geometric random variable (discrete-time case). So, the corresponding models become Markovian and have a useful max-autoregressive representation, i.e., 
\begin{equation}\label{model space-time}
X(\S,t)=\max\left\{\delta X(\boldsymbol{s}-\boldsymbol{\tau},t-1),(1-\delta)H(\boldsymbol{s},t)\right\}, \   (\S,t) \in \mathcal {S}\times \mathcal{T},
\end{equation}
where the parameter $\delta \in (0,1)$ measures the influence of the past, the parameter $\boldsymbol{\tau} \in \mathbb{R}^2$ represents some kind of specific direction of propagation/contagion in space and $H=:\{H(\S,t), \S \in \mathcal{S}, t \in \mathcal{T}\}$ is a time-independent process and is derived from independent replications of a spatial max-stable process $\{H(\S), \S \in \mathcal{S}\}$. This model can be seen as an extension of the real-valued max-autoregressive moving-average
process MARMA(1,0) to the spatial context, see \cite{davis1989basic}. The value at location $\boldsymbol{s}$ and time $t$ is either related to the value at location $\boldsymbol{s}-\boldsymbol{\tau}$ at time $t-1$ or to the value of another process (the innovation), $H$, that characterizes a new event happening at location $\boldsymbol{s}$. This model may be useful for phenomena that propagate in space.

In the following, we will focus on the processes satisfying (\ref{model space-time}). Let $V_{\boldsymbol{0},\boldsymbol{h}-l\boldsymbol{\tau}}$ denote the exponent function characterizing the spatial distribution of the process $H(\S,t)$, then the bivariate c.d.f. $F_{\H,l}$ of $(X(\boldsymbol{0},0),X( \H,l))$ can be expressed for $x_1,x_2>0$ as 
\begin{equation} \label{Biv CDF}
-\log F_{\H,l}(x_1,x_2)  =V_{\boldsymbol{0},\boldsymbol{h}-l\boldsymbol{\tau}} \left(x_1,\frac{x_2}{\delta^{l}}\right)+\frac{1-\delta^{l}}{x_2}.
\end{equation}
Moreover, the spatio-temporal extremal dependence function in (\ref{Def::spatiotemporal-theta}) can be easily deduced in this case by setting $x_1=x_2=x$ in (\ref{Biv CDF}),
\begin{equation}\label{spatiotemporal theta}
\theta(\H,l)=V_{\boldsymbol{0},\H-l\boldsymbol{\tau}}\left(1,\delta^{-l} \right) + 1-\delta^{l}.
\end{equation}

Clearly, space and time are not fully separated in the extremal dependence function, even if $\boldsymbol{\tau}=\boldsymbol{0}$ (space and time are completely separated in the spectral representation). Asymptotic time independence is achieved when $\lim_{l \, \to \, \infty} \theta(\H,l)\rightarrow 2$. In the sequel, we give two examples of a bivariate space-time max-stable process satisfying (\ref{model space-time}). 
\begin{enumerate}[label=(\roman*)]
	\item \textbf {\em Spectrally separable space-time max-stable Smith process}
	
If the innovation process $H$ is derived from independent replications of a spatial Smith process \cite{smith1990max} with a covariance matrix $\boldsymbol{\Sigma}$. Then the bivariate c.d.f. $F_{\H,l}$ of the resulting spatio-temporal model in (\ref{model space-time}) has the form    
	\begin{align} \label{Smith::ERWAN}
	-\log F_{\H,l}(x_1,x_2)& =\frac{1}{x_{1}}\Phi\left( \frac{b (\H,l)}{2}+\frac{1}{b(\H,l)}\log\left(\frac{x_{2}}{\delta^{l} x_{1}}\right)\right )\\ \nonumber&+ \frac{\delta^{l}}{x_{2}}\Phi\left( \frac{b(\H,l)}{2}+\frac{1}{b(\H,l)}\log\left(\frac{\delta^{l}x_{1}}{x_{2}}\right)\right)+\frac{1-\delta^{l}}{x_2},
	\end{align}
	where $b(\H,l)=\sqrt{(\boldsymbol{h}-l\boldsymbol{\tau})^{t}\boldsymbol{\Sigma}^{-1}(\boldsymbol{h}-l\boldsymbol{\tau})}$. The associated spatio-temporal extremal dependence function with this model is
	\begin{align}
	\theta(\H,l)=& \Phi\left( \frac{b(\H,l)}{2}+\frac{1}{b(\H,l)}\log\left(\delta^{-l}\right)\right)+ \delta^{l}\Phi\left( \frac{b(\H,l)}{2}+\frac{1}{b(\H,l)}\log\left(\delta^{l}\right)\right) \\ \nonumber & + 1-\delta^{l}.
	\end{align}
	\item \textbf{\em Spectrally separable space-time max-stable Schlather process}
	
	The spatio-temporal model in (\ref{model space-time}) with an innovation process $H$ derived from independent replications of a spatial Schlather process \cite{schlather2002models}, has a bivariate c.d.f. $F_{\H,l}$ of the form
	\begin{align} \label{Schlather space time}
	-\log F_{\H,l}(x_1,x_2) &=\frac{1}{2}\left( \frac{1}{x_1}+\frac{\delta^{l}}{x_2}\right)\\ \times \nonumber &\left[\left(1+ \sqrt{1- \frac{2\delta^l (\rho(\H,l)+1)x_{1} x_{2}} {(\delta^{l}x_{1}+x_{2})^{2}}}\right)\right] + \frac{1-\delta^{l}}{x_2},
	\end{align}
	where $\rho(\H,l)$ is the spatio-temporal exponential correlation function related to this model. The associated spatio-temporal extremal coefficient with this model is
	$\theta(\H,l)=\frac{1}{2} (1+\delta^{l})\left[\left(1+ \sqrt{1- \frac{2\delta^l (\rho(\H,l)+1)} {(1+\delta^{l})^{2}}}\right)\right]+ 1-\delta^{l}.$

\end{enumerate}
If the time lag $l=0$, the formulas in (\ref{Smith::ERWAN}) and (\ref{Schlather space time}) reduce to the bivariate distributions of the max-stable spatial fields.

	\section{Statistical inference for space-time max-stable processes }  \label{sec:inference}
In what follows, we shall denote, respectively, by $h=\HH=: \lVert \S_1-\S_2 \rVert, \ \H \in \R^2$ and $l'=|l|=:|t_1-t_2|, \ l \in \R$ the Euclidean norm of spatial lag $\H$ and the absolute value of temporal lag $l$. 

We now describe two semiparametric estimation schemes for space-time max-stable processes based on the spatio-temporal $F$-madogram in (\ref{Def::spatiotemporal-mado}), which stems from a classical geostatistical tool; the madogram \cite{matheron1987suffit}. It has a clear link with extreme value theory throughout the spatio-temporal extremal dependence function $\theta(.)$, i.e., 
\begin{equation}
\nu_{F}({\H},l)=\frac{1}{2} -\frac{1}{\theta(\H,l)+1}.
\label{spatiotemporal madogram}
\end{equation}

In practice, measurements are typically taken at various locations, sometimes on a grid, and at regularly spaced time intervals. In the following, the process $X:=\{X(\S,t): {(\S,t)\in \mathcal{S}\times \mathcal{T}} \}$ is assumed to be a stationary   space-time max-stable process. It is observed on locations assumed to lie on a regular 2-dimensional (2D) grid, i.e.,
$$ S_n = \left\{ \S_{i}: i=1,\ldots, n^2 \right\} =  \left\{ (x,y),\ x,y \in \left\{1,\ldots,n\right\}\right\} ,$$
and at equidistant time moments, given by $\{t_1,\ldots,t_T\}=\{1,\ldots,T\}$. This sampling scheme has been adopted in various studies in the literature, see e.g., \cite{davis2013statistical,buhl2016anisotropic,buhl2016semiparametric}.  For statistical inference on the process $X$, we develop the following two semiparametric estimation schemes.

\subsection{Scheme 1} \label{Sec::Scheme1}

Let $\boldsymbol{\psi} =( {\boldsymbol{\psi}}^{(s)}, {\boldsymbol{\psi}}^{(t)})$ denotes the vector gathering the parameters of the process $X$ to be estimated, where ${\boldsymbol{\psi}}^{(s)}$ and ${\boldsymbol{\psi}}^{(t)}$ denote, respectively, the vectors gathering the spatial and temporal parameters. In this scheme, we consider how the process evolves at given time of reference (a merely spatial process), and its evolution over time at a given location (a merely temporal process). So, ${\boldsymbol{\psi}}^{(s)}$ and ${\boldsymbol{\psi}}^{(t)}$ can be estimated separately. More precisely, denote by $\mathcal{H} \subset [0,\infty)$ and $\mathcal{K} \subset [0,\infty)$ finite sets of spatial and temporal lags on which the estimation is performed. Let the set $\mathcal{B}_h$ summarizes all pairs of $ \mathcal{S}_n$ which give rise to the same spatial lag $h \in \mathcal{H}$, i.e.,
$$ \mathcal{B}_{h} =\{(\ell,p) \in \{1,\ldots,n^2\}^2: \lVert \boldsymbol{s}_\ell-\boldsymbol{s}_p\rVert = \HH= h\}.$$


The inferential methodology is summarized in the following steps:

\begin{enumerate}[label=(\roman*)]
	\item 	As a first step, we estimate the purely spatial/temporal $F$-madogram nonparametrically by the empirical version. Denote by ${\widehat{\nu}}^{(t)}_{F} (\boldsymbol{h}), \ \HH \in \mathcal{H}$ $\left( \text{respectively} \ {\widehat{\nu}}^{(\boldsymbol{\S})}_{F} (l'), \ l' \in \mathcal{K}\right)$ the nonparametric estimate of the purely spatial (respectively temporal) $F$-madogram. As is standard in geostatistics, we compute ${\widehat{\nu}}^{(t)}_{F} (\boldsymbol{h})$ from the empirical spatio-temporal $F$-madogram ${\widehat{\nu}}_F(\H,l)$ at spatio-temporal distances $(\H,0)$, that is for all $\{t_1,\ldots,t_T\}$,
	$${\widehat{\nu}}^{(t)}_{F} (\boldsymbol{h})={\widehat{\nu}}_F(\H,0)= \frac{1}{2 |\mathcal{B}_{h}|} \underset{\lVert \boldsymbol{s}_\ell-\boldsymbol{s}_p\rVert = \lVert \boldsymbol{h}\rVert= h}{\sum_{p=1}^{n^2}\sum_{\ell=1}^{n^2}} |F\{X (\boldsymbol{s}_\ell,t)\} - F\{X(\boldsymbol{s}_p,t)\}|\/,\ {h}  \in \mathcal{H},$$ 
	where $|.|$ denotes the cardinality of the set $B_{h}$ and $F$ is the standard Fr\'{e}chet probability distribution function. Let us remark that, a similar estimator in the framework of $\lambda$-madogram has been adopted by \cite{naveau2009modelling} in an analysis of Bourgogne (France) annual maxima of daily rainfall measurements. On the other hand, ${\widehat{\nu}}^{(\boldsymbol{s})}_{F} (l')$ is computed from the empirical spatio-temporal $F$-madogram ${\widehat{\nu}}_F(\H,l')$ at spatio-temporal distances $(\boldsymbol{0},l')$, that is for all $\boldsymbol{s} \in   \mathcal{S}_n $
	$${\widehat{\nu}}^{(\boldsymbol{s})}_{F} (l')={\widehat{\nu}}_F(\boldsymbol{0},l')= \frac{1}{2 (T-l')} \sum_{k=1}^{T-l'} |F\{X (\S,t_k)\} - F\{X(\S,t_{k +l'})\}|\/, \ l' \in \mathcal{K}.$$

	\item	Then, the overall purely spatial (respectively temporal) $F$-madogram estimates ${\widehat{\nu}}_{F} (\boldsymbol{h})$ (respectively ${\widehat{\nu}}_{F} (l')$) are computed from the means over the temporal moments (respectively the spatial locations). More precisely,
	\begin{equation} \label{Spatial F}
	{\widehat{\nu}}_{F} (\boldsymbol{h})= \frac{1}{T} \underset{\lVert \boldsymbol{h}\rVert = h }{\sum_{k=1}^{T}}{\widehat{\nu}}^{(t_k)}_{F}(\boldsymbol{h}), \ h \in \mathcal{H}.
	\end{equation}
	\begin{equation} \label{Temporal F}
	{\widehat{\nu}}_{F} (l')=  \frac{1}{n^2}\sum_{\ell=1}^{n^2} {\widehat{\nu}}^{(\boldsymbol{s}_\ell)}_{F} (l'), \ l' \in \mathcal{K}. 
	\end{equation}
	\item Finally, a NLS procedure is applied to estimate the parameters of interest.
	\begin{equation} \label{eq:psi1}
	\boldsymbol{{\widehat{\boldsymbol{\psi}}}}^{(s)}=\underset{ {\boldsymbol{\psi}}^{(\S)} \in {\boldsymbol{\Psi}}^{(\S)}} {\text{arg min}}  \sum_{\Vert\boldsymbol{h}\Vert =h \in \mathcal{H}} \omega^{\H}  \left({\widehat{\nu}}_{F} (\boldsymbol{h})- \nu_{F}^{(\S)} (\boldsymbol{h},\boldsymbol{\psi}^{(s)})
	\right)^{2}, \ h \in \mathcal{H},	
	\end{equation}
	\begin{equation} \label{eq:psi2}
	\boldsymbol{{\widehat{\boldsymbol{\psi}}}}^{{(t)}}=\underset{ {\boldsymbol{\psi}}^{{(t)}} \in {\boldsymbol{\Psi}}^{{(t)}} } {\text{arg min}}  \sum_{l' \in \mathcal{K}} \omega^{l'} \left({\widehat{\nu}}_{F} (l')- \nu_{F}^{(t)} (l',\boldsymbol{\psi}^{{(t)}})
	\right)^{2}, \ l'\in \mathcal{K},	
	\end{equation}
	where $\nu_{F}^{(\S)} (\boldsymbol{h},\boldsymbol{\psi}^{(s)})=\nu_{F} (\H,0,\boldsymbol{\psi}^{{(s)}})$ and $\nu_{F}^{(t)}  (l',\boldsymbol{\psi}^{{(t)}})=\nu_{F} (\boldsymbol{0},l',\boldsymbol{\psi}^{{(t)}})$ denote, respectively, the spatial and temporal model-based $F$-madogram counterparts. $\omega^{\H} \geq 0 $ and $\omega^{l'} \geq 0$ denote, respectively, the spatial and temporal weights. Since it is expected that the spatio-temporal pairs which are far away in space or in time, have only little influence on the dependence parameters to be estimated, a simple choice for these weights is $\omega^{\H}= 
	\mathbbm{1}_{\{ \HH \leq r\}}$, $\omega^{l'}= 
	\mathbbm{1}_{\{ l' \leq q\}}$, where $\mathbbm{1}(.)$ denotes the indicator function and $(r,q)$ is fixed.	
\end{enumerate}
Note that the setup of the inferential methodology in Scheme 1 is close to the one proposed in \cite{buhl2016semiparametric}, in which the spatio-temporal extremogram in (\ref{Space-time::Extremo}) was adopted.


\subsection{Scheme 2}

We now generalize Scheme 1 in order to estimate temporal and spatial parameters simultaneously. Thus, we consider how the process $X$ evolves in both space and time. In the classical geostatistics, for a stationary spatio-temporal process $\left\{ X(\S,t): (\S,t) \in \mathcal{S}\times \mathcal{T} \right\}$, the spatio-temporal empirical classical semivariogram is defined by
$$\widehat{\gamma}(\H,l)= \frac{1}{2 \lvert \mathcal{B}_{(\H,l)}\rvert} \sum_{\mathcal{B}_{(\H,l)}} \left(X(\S_i,t_i,) - X(\S_j,t_j)\right)^2,$$
where $\mathcal{B}_{(\H,l)}=\left\{  (\S_i,t_i)(\S_j,t_j): \S_i-\S_j=\H \ \text{and} \ t_i-t_j=l  \right\}$, see e.g., \cite{fernandez2015spatial}. By adapting this estimator to our framework, we consider the following estimation procedure:
\begin{enumerate}[label=(\roman*)]
	\item 	First, the spatio-temporal $F$-madogram is estimated nonparametrically by its empirical version. Assume the set $\mathcal{B}_{(h,l')}$ summarizes all pairs of $ \mathcal{S}_n$ which give rise to the same spatial lag $h \in \mathcal{H} \subset [0,\infty)$ and the same temporal lag $l' \in \mathcal{K} \subset [0,\infty)$.  In other words, combining the spatial and the temporal lags from Scheme 1, i.e.,
	$$ \mathcal{B}_{(h,l')} =\left\{\left(\S_i,t_i),(\boldsymbol{s}_j,t_j)\right): \lVert \boldsymbol{s}_i-\boldsymbol{s}_j\rVert = h, |t_i-t_j|=l' \right\}.$$
	We estimate $\nu_F(\H,l')$ by
	\begin{equation}
	{\widehat{\nu}}_F(\H,l')= \frac{1}{2 |\mathcal{B}_{(h,l')}|}   {{\sum_{ \mathcal{B}_{(h,l')}}}} |F\{X (\boldsymbol{s}_i,t_i)\ - F\{X(\boldsymbol{s}_j,t_j)\}|\/,
	\end{equation}
	where $|.|$ denotes the cardinality of the set $\mathcal{B}_{(h,l')}$ and $(h,l') \in  \mathcal{H} \times \mathcal{K}$. 
	
	\item Then, we apply a NLS fitting to obtain the estimates of the process parameters; $\boldsymbol{\psi}$, i.e., 
	\begin{equation} \label{eq:psi}
	\boldsymbol{{\widehat{\boldsymbol{\psi}}}}=\underset{{\boldsymbol{\psi}} \in {\boldsymbol{\Psi}}} {\text{arg min}} \underset{ \lVert\boldsymbol{h} \rVert=h} {\sum_{l' \in \mathcal{K}} \sum_{h \in \mathcal{H}}} \omega^{\H,l'}  \left({\widehat{\nu}}_{F} (\H,l')- \nu_{F} (\H,l',\boldsymbol{\psi})
	\right)^{2}, \ (h,l') \in  \mathcal{H} \times \mathcal{K},	
	\end{equation}
	where $\omega^{\H,l'} \geq 0$ denotes the spatio-temporal weights and $\nu_{F} (\boldsymbol{h},l',\boldsymbol{\psi})$ is the model-based spatio-temporal $F$-madogram.
\end{enumerate}
The idea underlying the construction of Scheme 2 is that when modeling and predicting a given phenomenon, significant benefits may be obtained by considering how it evolves in both space and time rather than only considering its spatial distribution at a given time of reference (a merely spatial process), or its evolution over time at a given location (a merely temporal process), such as those described in Scheme 1. Lastly, the establishment of the asymptotic properties of the resulting pairwise dependence estimates is deferred to future work. The derived asymptotic properties of the unbinned empirical $\lambda$-madogram in the spatial context, see \cite{naveau2009modelling} (Proposition 3 and 4) might provide a starting point. Nevertheless, this setting is more specialized. In the real data example of that study, a binned version of the empirical $\lambda$-madogram is adopted and deriving the convergence of this estimator as the cardinality of the distance class (i.e., $\mathcal{B}_h$) increases is still challenging. Therefore, we will provide some numerical indications for the asymptotic properties of our pairwise dependence estimates.

\subsection{Illustration examples} 
In order to illustrate how the proposed estimation schemes perform, we consider the following two examples, which we will revisit in Section~\ref{sec:simulation}.
\begin{ex} \label{example1}
	\textbf{(Estimation of isotropic space-time max-stable BR)}
	Let us consider the space-time max-stable BR process in (\ref{BR spectral fun}) with bivariate c.d.f. (\ref{Biv BR}), where the dependence structure is given by the following stationary isotropic fractional Brownian motion (FBM) spatio-temporal semivariogram 
	\begin{equation}
	\gamma(\H,l) :=\gamma(h,l')=  2 \phi_s h^{\kappa_s}+2 \phi_t {l'}^{\kappa_t} , 
	\label{gamma}
	\end{equation}
	where the scalar distance $h=\HH=\lVert \S_1-\S_2 \rVert$, $l' = \lvert l \rvert=\lvert t_1-t_2 \rvert$, $ \phi_s,  \phi_t >0$ determine spatial and temporal scale parameters and $\kappa_s,\kappa_t \in (0,2]$ relate to the smoothness of the underlying Gaussian process in space and time. The associated spatio-temporal $F$-madogram with this process is 
	\begin{equation} \label{BR::F-mado}
	\nu_F(h,l')= \frac{1}{2} - \frac{1}{2\Phi\left(\sqrt{\phi_s h^{\kappa_s} + \phi_t l'^{\kappa_t}}\right)+1},
	\end{equation}
	where $\theta(h,l')=2\Phi\left(\sqrt{\phi_s h^{\kappa_s} + \phi_t l'^{\kappa_t}}\right)$ is the associated spatio-temporal extremal dependence function. Figure~\ref{3DBR} visualizes a 3D representation of the spatio-temporal FBM semivariogram in (\ref{gamma}) and the associated dependence summary measures: the spatio-temporal extremal dependence function $\theta: \mathbb{R}^{2} \times \mathbb{R}^{+}  \mapsto[1,2]$ and the spatio-temporal $F$-madogram $\nu_F: \mathbb{R}^{2} \times \mathbb{R}^{+}  \mapsto[0,1/6]$. Complete dependence (respectively complete independence) is achieved at lower boundaries (respectively upper boundaries). Moreover, Figure~\ref{Theoreticalbehavior} displays the theoretical behaviors of the purely spatial FBM semivariogram $\gamma^{\boldsymbol{(\S)}}(h,\kappa_s)$ and the related purely spatial $F$-madogram $ \nu_{F}^{\boldsymbol{(\S)}} (h,\kappa_s)$. Obviously, depending on the value of the smoothness parameter $\kappa_s$, these measures exhibit a large variety of dependence behaviors. 
	\begin{figure} [!h]
		\centering
		\includegraphics[scale=0.4]{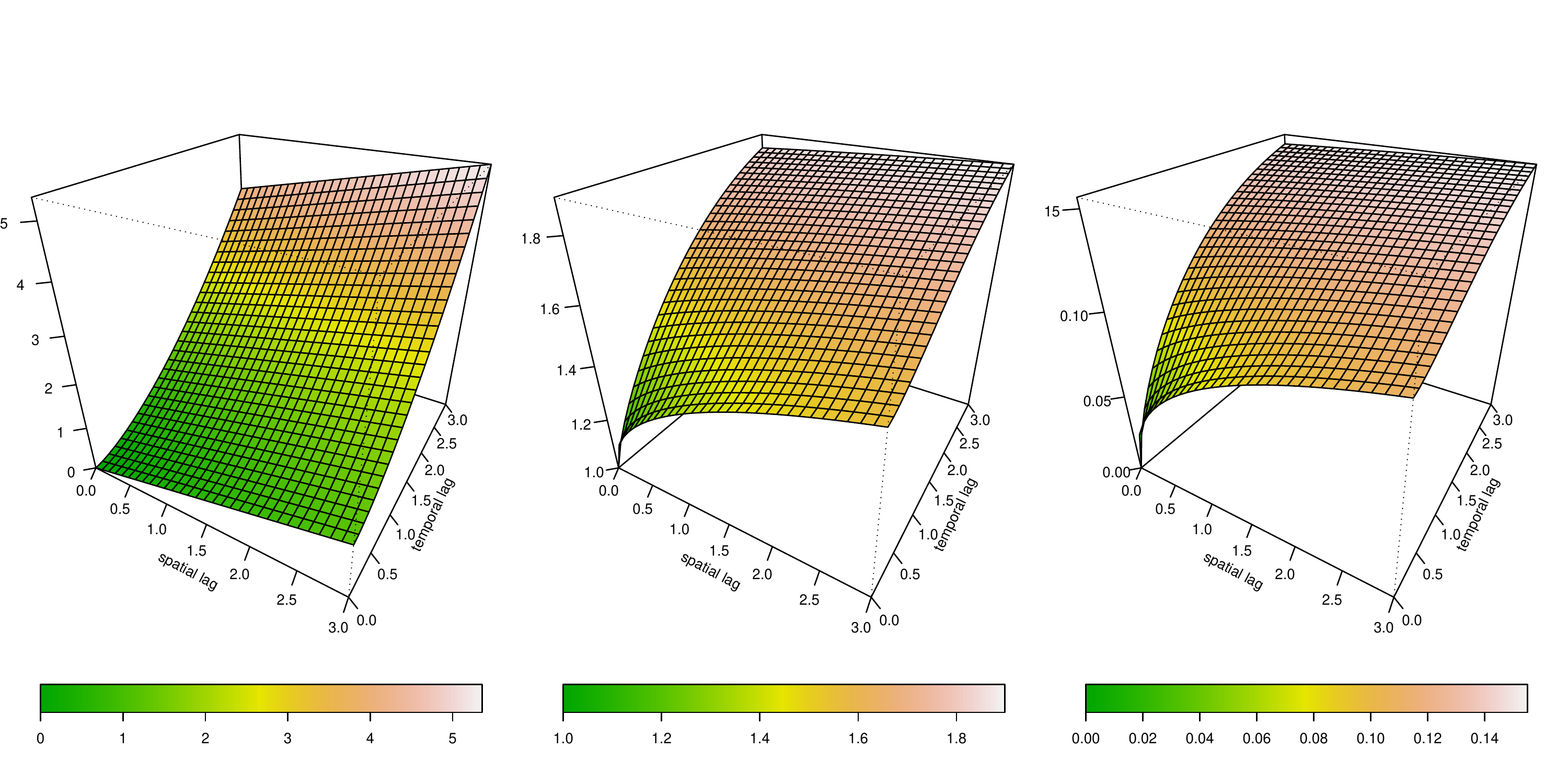}
		
		\caption{Spatio-temporal FBM semivariogram $\gamma(h,l')=0.8 h^{1.5}+0.4l'$ (left panel). The associated spatio-temporal extremal dependence function (middle panel). The associated spatio-temporal $F$-madogram (right panel).}
		
		\label{3DBR}
		
	\end{figure}

	\begin{figure} [!h]
		\centering
		\includegraphics[scale=0.48]{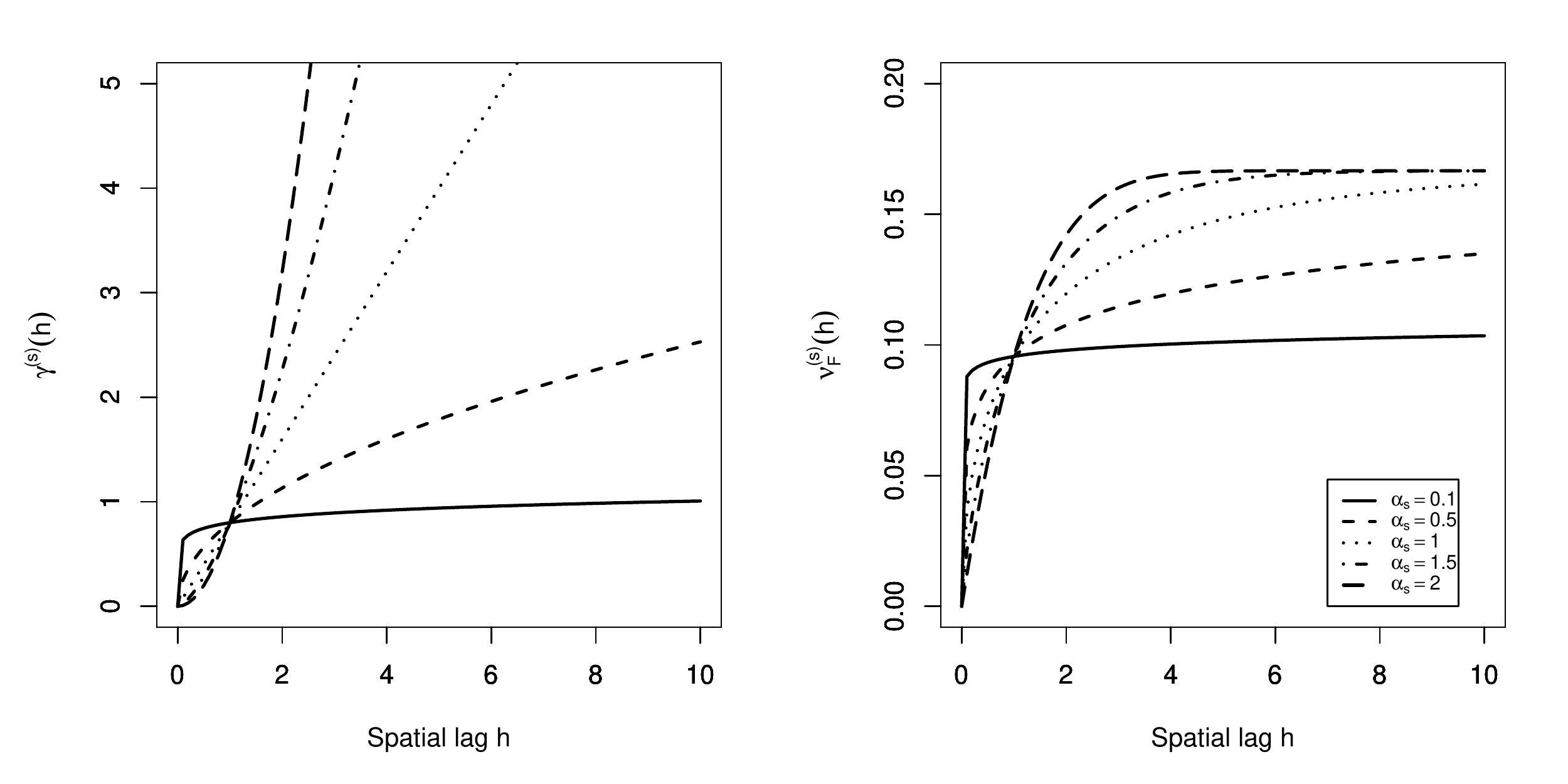}
		
		\caption{The FBM semivariogram $\gamma^{(\S)}(h,\kappa_s)= 0.8 h^{\kappa_s}$ (left panel) and the related spatial $F$-madogram $ \nu_{F}^{(\S)} (h,\kappa_s)=0.5 - \left\{2\Phi\left(\sqrt{ 0.4 h^{\kappa_s}}\right)+1\right\}^{-1}$ (right panel) plotted as functions of space lag $h$, with different smoothness parameter $\kappa_s \in \{0.1,0.5,1,1.5,2\}$. }
		
		\label{Theoreticalbehavior}
		
	\end{figure}
	
	With this construction, based on Scheme 1, the NLS optimization problems in (\ref{eq:psi1}) and (\ref{eq:psi2}) can be expressed as
	\begin{equation} \label{eq:BR1}
	\begin{pmatrix}
	{\widehat{\kappa}}_s\\ {\widehat{\phi}}_s  
	\end{pmatrix}=\underset{{\substack{\\\phi_s>0 \\ \kappa_s \in (0,2]}}}  {\text{arg min}}  \sum_{h \in \mathcal{H}} \omega^h \left({\widehat{\nu}}_{F} (h)- \left\{\frac{1}{2} - \frac{1}{2\Phi\left(\sqrt{\phi_s h^{\kappa_s}}\right)+1} \right\} 
	\right)^{2}, \ h \in \mathcal{H},	
	\end{equation}
\begin{equation} \label{eq:BR2}
	\begin{pmatrix}
	{\widehat{\kappa}}_t\\ {\widehat{\phi}}_t  
	\end{pmatrix}= \underset{{\substack{\\\phi_t>0 \\ \kappa_t \in (0,2]}}} {\text{arg min}} \sum_{l' \in \mathcal{K}} \omega^{l'} \left({\widehat{\nu}}_{F} (l')- \left\{\frac{1}{2} - \frac{1}{2\Phi\left(\sqrt{\phi_t l'^{\kappa_t}}\right)+1} \right\}  
	\right)^{2}, \ l' \in \mathcal{K}.	
	\end{equation}
	
	Lastly, with $(h,l') \in  \mathcal{H} \times \mathcal{K}$ and on the basis of Scheme 2, the NLS estimation problem in ({\ref{eq:psi}}) has the form
	\begin{equation} \label{eq:BR}
	\begin{pmatrix}
	{\widehat{\kappa}}_s\\ {\widehat{\phi}}_s \\{\widehat{\kappa}}_t\\ {\widehat{\phi}}_t
	\end{pmatrix}=\underset{{\substack{\\\phi_s,\phi_t>0 \\ \kappa_s,\kappa_t \in (0,2]}}}  {\text{arg min}} \sum_{l' \in \mathcal{K}}  \sum_{h \in \mathcal{H}} \omega^{h,l'} \left({\widehat{\nu}}_{F} (h,l')- \left\{\frac{1}{2} - \frac{1}{2\Phi\left(\sqrt{\phi_s h^{\kappa_s}+\phi_t {l'}^{\kappa_t} }\right)+1} \right\} 
	\right)^{2}.	
	\end{equation}
\end{ex}

\begin{ex} \label{example2}
	\textbf{(Estimation of spectrally separable space-time max-stable Smith process)}
	We now describe the way to fit the spectrally separable space-time max-stable Smith process. Indeed, the estimation procedure can be simplified since the purely spatial parameters can be estimated independently of the purely temporal parameters. Formally, we consider the process in (\ref{model space-time}), where the innovation process $H$ is derived from independent replications of a spatial Smith process with covariance matrix
	\begin{equation} \label{Sigma}
	\boldsymbol{\Sigma}=\begin{pmatrix}
	\sigma_{11}&\sigma_{12} \\
	\sigma_{12} &\sigma_{22}
	\end{pmatrix}. 
	\end{equation}
	
	We donte by $\boldsymbol{\psi}$ the vector gathering the parameters to be estimated, i.e., $ \boldsymbol{\psi}=\left(\sigma_{11}, \sigma_{12}, \sigma_{22}, \boldsymbol{\tau}^{t},\delta \right)^{t}$. It is possible to separate the estimation. Firstly, the estimation of the spatial parameters ${\boldsymbol{\psi}}^{({s})}=\left(\sigma_{11}, \sigma_{12},\sigma_{22}\right)^{t}$ is carried out. Secondly, once ${\boldsymbol{\psi}}^{({s})}$ is known, it is held fixed and we estimate the temporal parameters ${\boldsymbol{\psi}}^{(t)}=\left( \boldsymbol{\tau}^{t},\delta\right)^{t} = \left( \tau_1, \tau_2, \delta\right)^t$. Subsequently, under Scheme 1, the NLS optimization problems in (\ref{eq:psi1}) and (\ref{eq:psi2}) can be expressed as
	\begin{equation} \label{eq:Smith1}
	\begin{pmatrix}
	{\widehat{\sigma}}_{11}\\ {\widehat{\sigma}}_{12}  \\ {\widehat{\sigma}}_{22} 
	\end{pmatrix}= \underset{{\substack{\\ \sigma_{11},\sigma_{22} >0 \\ \sigma_{12} \in \R }}} {\text{arg min}} \underset{\lVert\boldsymbol{h}\rVert=h} {\sum_{h\in \mathcal{H}}} \omega^{\H} \left({\widehat{\nu}}_{F} (\boldsymbol{h})- \left\{\frac{1}{2} - \frac{1}{2\Phi\left(\sqrt{\boldsymbol{h}^{t}\boldsymbol{\Sigma}^{-1}\boldsymbol{h}}/2\right)+1} \right\} 
	\right)^{2}, \ {h} \in \mathcal{H},	
	\end{equation}
	\begin{equation} \label{eq:Smith2}
	\begin{pmatrix}
	{\widehat{\delta}}\\ {\widehat{\tau}}_1\\ {\widehat{\tau}}_2
	\end{pmatrix}= \underset{{\substack{\\ a \in (0,1)\\ \tau_1,\tau_2 \in \R }}} {\text{arg min}} \sum_{l' \in \mathcal{K}} \omega^{l'}\left({\widehat{\nu}}_{F} (l')- \left\{\frac{1}{2} - \frac{1}{\theta(l')+1} \right\}  
	\right)^{2}, \ l' \in \mathcal{K},	
	\end{equation}
	where, 
	$$ \theta(l')= \Phi\left( \frac{b^{*}(l')}{2}+\frac{1}{b^{*} (l')}\log\left(\delta^{-l'}\right)\right)+ \delta^{l}\Phi\left( \frac{b^{*}(l')}{2}+\frac{1}{b^{*}(l')}\log\left(\delta^{l'}\right)\right)+ 1-\delta^{l'} $$
	with $b^{*}(l')=\sqrt{( \boldsymbol{0}-l'  \boldsymbol{\tau})^{t}\widehat{\boldsymbol{\Sigma}}^{-1} ( \boldsymbol{0}-l' \boldsymbol{\tau}) }.$
	
	In order to figure out the role of the temporal parameter $\delta$ for this process. For a fixed site $\boldsymbol{s} \in \mathcal{S}$, Figure~\ref{temporaltheta} displays the temporal extremal function $\theta(l')$ and the associated temporal $F$-madogram $ \nu_{F}^{(t)}$ for $\delta \in \{0.1,0.3,0.5,0.7,0.9\}$. We set $\boldsymbol{\Sigma}=$ 10 Id$_2$ and $\boldsymbol{\tau}=(1,1)^{t}$ (translation to the top right). Clearly, as the value of $\delta$ increases, the independece (i.e., $\theta(l')\rightarrow2$) occurs at larger time lags $l'$.
	\begin{figure} [!htp]
		\centering
		\includegraphics[scale=0.48]{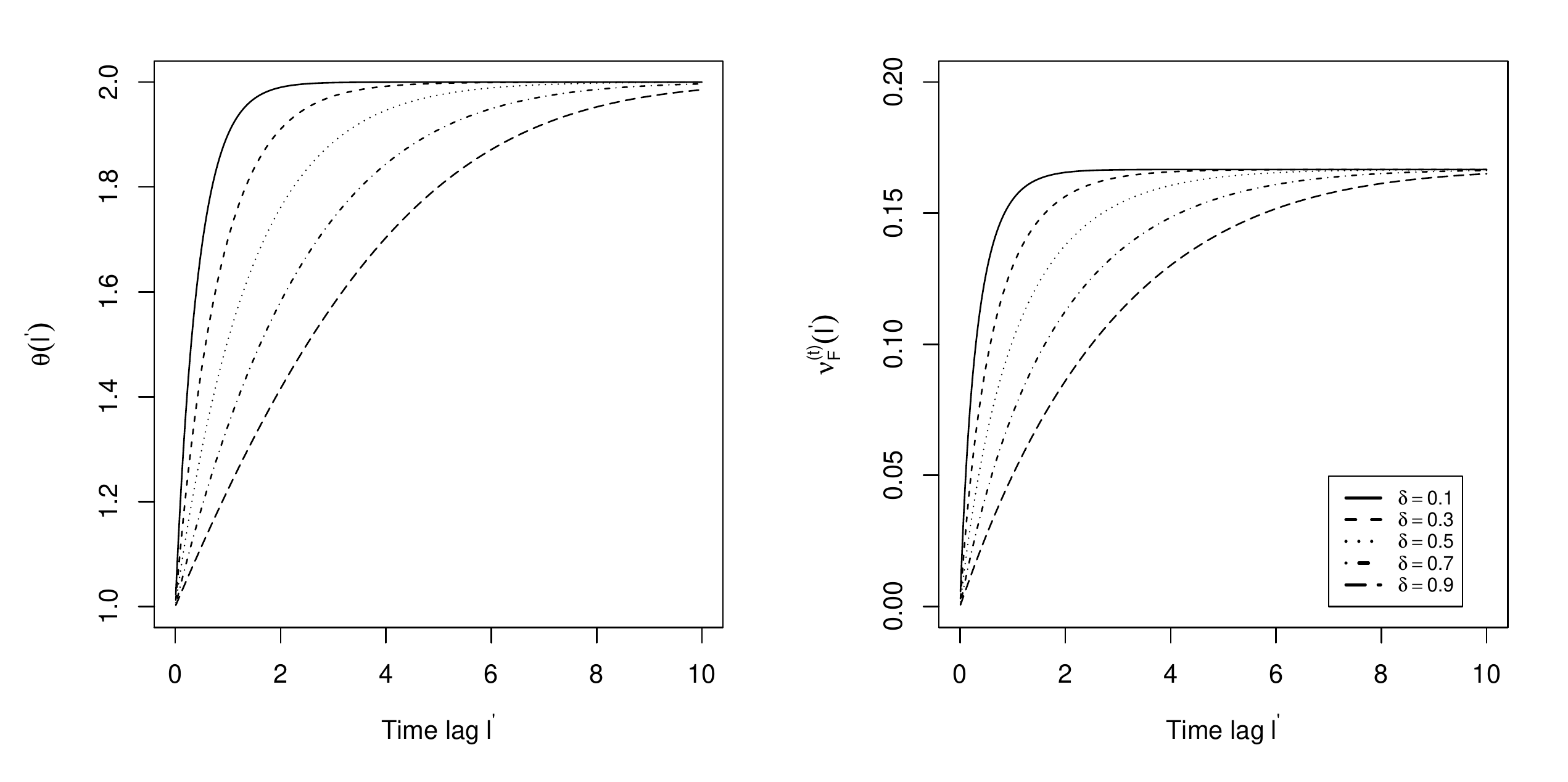}
		
		\caption{$\theta(l')$ and the associated $\nu_{F}^{(t)}(l')$ plotted as functions of time lag $l'$ for $\delta \in \{0.1,0.3,0.5,0.7,0.9\}$ based on the process (\ref{model space-time}), where 
			$H$ is a sequence of i.i.d. spatial Smith processes with covariance matrix $\boldsymbol{\Sigma}=$ 10 Id$_2$. }
		
		\label{temporaltheta}
		
	\end{figure}

	Lastly, based on Scheme 2, the NLS estimator $\boldsymbol{\widehat{\psi}} = \left( \widehat{\sigma}_{11},\widehat{\sigma}_{12},\widehat{\sigma}_{22},\widehat{\tau}_1,\widehat{\tau}_2,\widehat{\delta}\right)^t$ is given by
	\begin{equation} \label{eq:Smith}
	{\widehat{\boldsymbol{\psi}}}= \underset{{\substack{\\ \boldsymbol{\psi} \in  \boldsymbol{\Psi}}}} {\text{arg min}} \underset{{\substack{\\ \lVert\boldsymbol{h}\rVert = h}}}{\sum_{l' \in \mathcal{K}} \sum_{h \in \mathcal{H}}}\omega^{\H,l'} \left({\widehat{\nu}}_{F} (\H,l' )- \left\{\frac{1}{2} - \frac{1}{\theta(\H,l' )+1} \right\}  
	\right)^{2}, \ (h,l') \in  \mathcal{H} \times \mathcal{K},	
	\end{equation}
	where
	$\theta(\H,l')= \Phi\left( \frac{b(\H,l')}{2}+\frac{1}{b(\H,l')}\log\left(\delta^{-l'}\right)\right)+ \delta^{l'}\Phi\left( \frac{b(\H,l')}{2}+ \frac{1}{b(\H,l')}\log\left(\delta^{l'}\right)\right)+ 1-\delta^{l'}$ $\text{with} \ b(\H,l')=\sqrt{({\boldsymbol{h}-l'\boldsymbol{\tau})}^{t}{\boldsymbol{\Sigma}}^{-1}{(\boldsymbol{h}-l'\boldsymbol{\tau}})}.$
\end{ex}

	
\section{Simulation study} \label{sec:simulation}

Throughout this section, we investigate the performance of the semiparametric estimation procedures introduced in Section~\ref{sec:inference} with three simulation studies.

\subsection{Simulation study 1: Fitting space-time max-stable BR process} \label{Sec::sim1}
In this study, we adopt the same experiment plan that has been proposed in \cite{buhl2016semiparametric} (Section~5), in order to make the results obtained there comparable with the results here.
\subsubsection{Setup for a simulation study} \label{Sec::setup1}
\begin{figure} [!htp]
	\centering
	\includegraphics[scale=0.4]{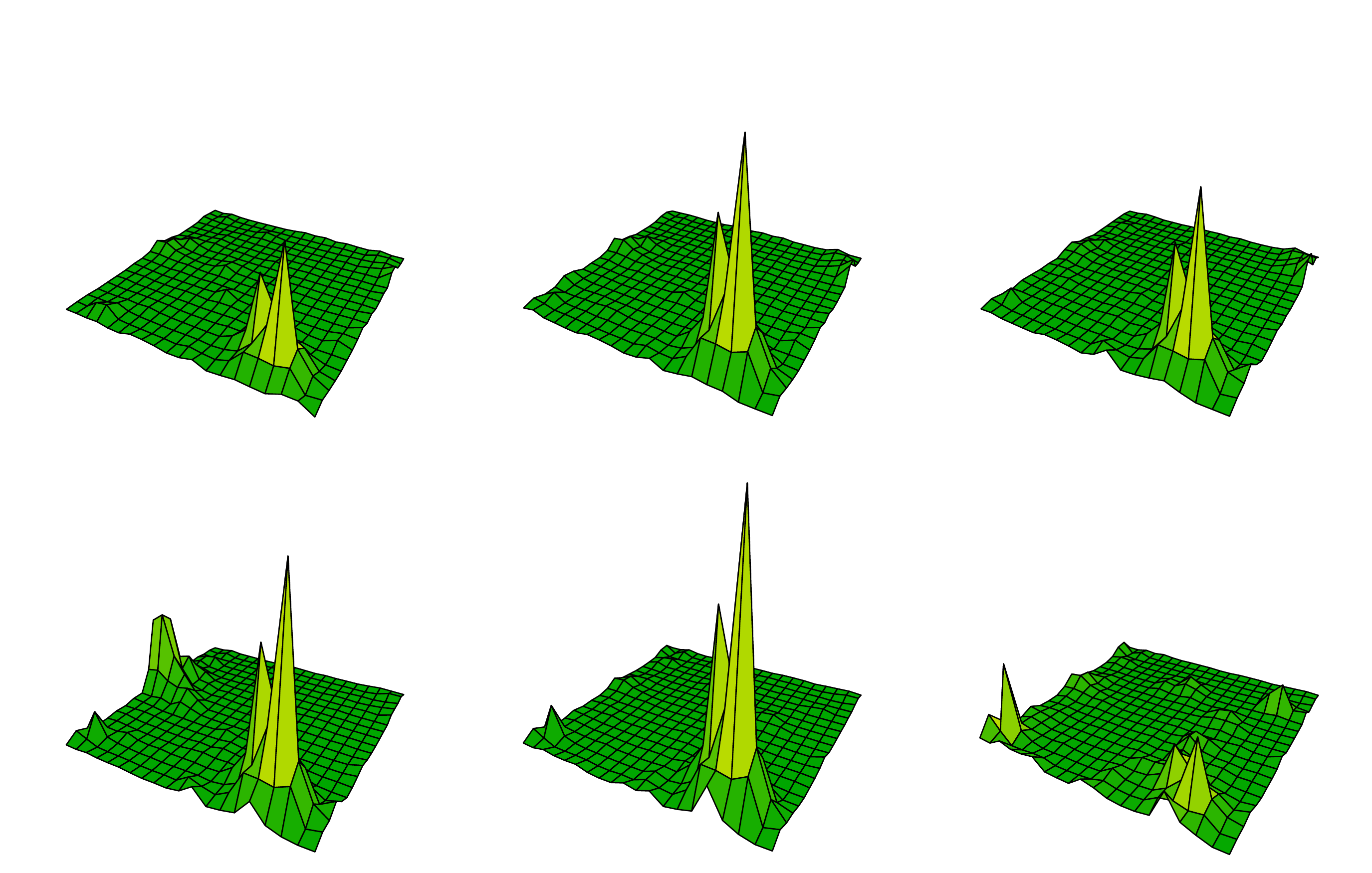}
	
	\caption{Simulation from a space-time max-stable BR process with spatio-temporal FBM semivariogram $\gamma(\H,l)= 0.8 h^{1.5}+0.4l'$ at six consecutive
		time points (from left to right and top to bottom).}
	
	\label{3DBRsim}
	
\end{figure}

\begin{figure} [htp!]
	\centering
	\includegraphics[scale=0.3]{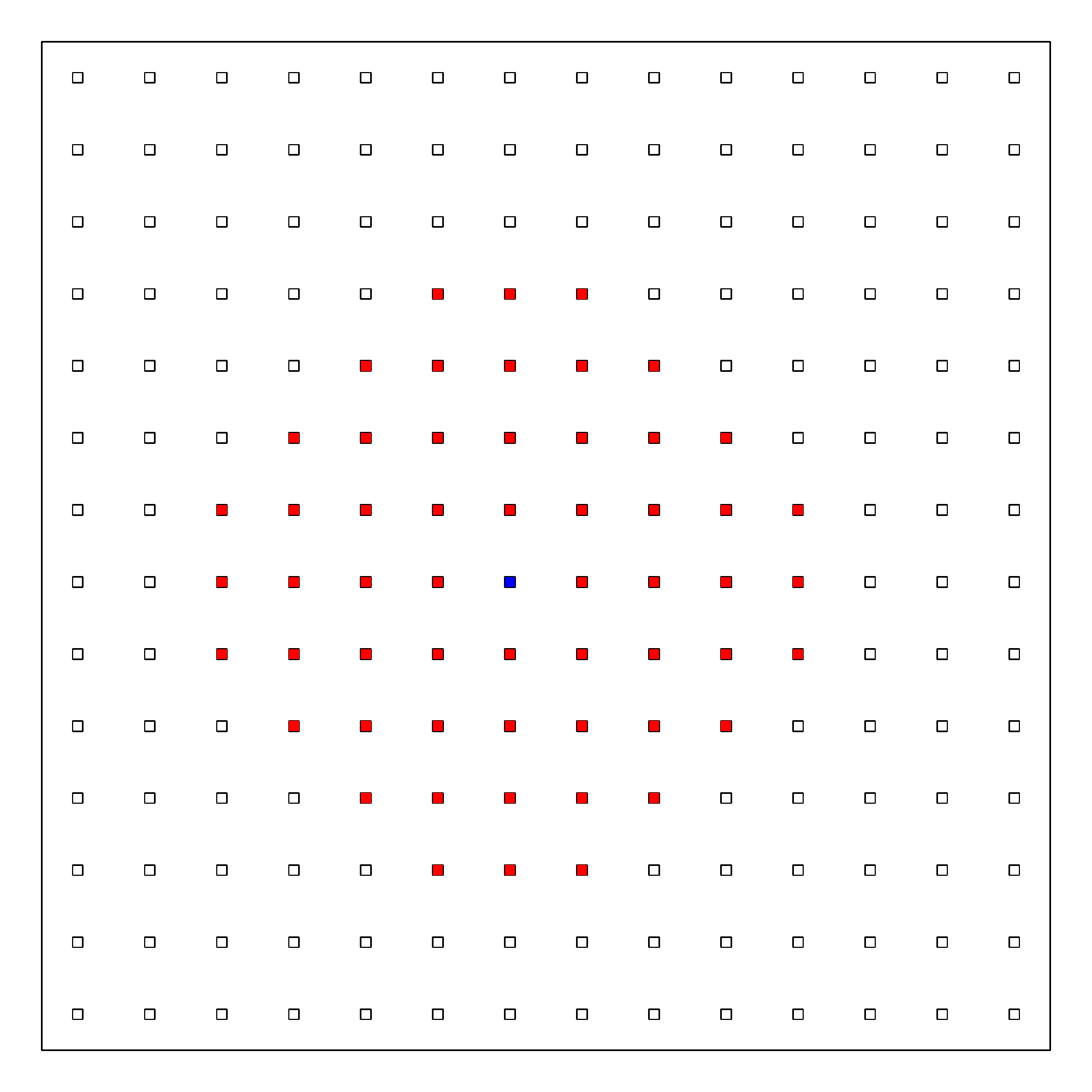}
	
	\caption{A regular $14 \times 14$ spatial grid. The distances between the peripheral locations (shown by red square symbols) and the central one (shown by blue square symbol) belong to the set $\mathcal{H}$. } 
	
	\label{GRIDSPATIAL}
	
\end{figure}

We simualte the space-time BR process with spectral representation (\ref{BR spectral fun}) and dependence function $\gamma$ modeled as in (\ref{gamma}). Namely,
\raggedbottom
\begin{equation} \label{FBM semi}
\gamma(\H,l) =  0.8 h^{3/2} +0.4 l'. 
\end{equation}
The simulations have been carried out using the function {\em RFsimulate} of the R package RandomFields \cite{schlatherRF} and based on the exact method proposed by \cite{dombry2016exact}. The space-time observation area is assumed to be on a $n \times n$ spatial grid and the time moments are equidistantly, i.e., $$ {\mathscr{A} } = \{(x,y):  x,y \in\{1,\ldots,n\} \} \times \{1,\ldots,T\}.$$ Figure~\ref{3DBRsim} visualizes a realization simulated from space-time BR process with a spatio-temporal FBM semivariogram model (\ref{FBM semi}) at six consecutive time points.
As in \cite{buhl2016semiparametric}, we choose the sets $\mathcal{H} =\{1,\sqrt{2},2,\sqrt{5},\sqrt{8},3,\sqrt{10},\sqrt{13},4,\sqrt{17}\}$ and $\mathcal{K}=\{1,\ldots,10\}$, where permutation tests show that these lags are enough to capture the relevant extremal dependence structure, see Figure~\ref{GRIDSPATIAL}. Equal weights are assumed. We repeat this experiment 100 times to obtain summary plots of the resulting estimates and to compute performance metrics: the mean estimate, the root mean squared error (RMSE) and the mean absolute error (MAE).

\subsubsection{Estimation using Scheme 1} \label{Sec::scheme1.1}

Simulation of space-time max-stable BR processes based on the exact method
proposed in \cite{dombry2016exact} can be time-consuming. Hence, for the sake of time-saving and due to the fact that the estimation of the purely spatial (respectively purely temporal) parameters depends on a large number of spatial observations (respectively a large number of observed time instants), we examine the performance of the purely spatial  (respectively purely temporal) estimates using two different space-time observation areas, i.e.,
\begin{itemize}
	\item {$ {\mathscr{A} }_1 = \{(x,y):  x,y \in\{1,\ldots,50\} \} \times \{1,\ldots,10\}.$}
	\item {$ {\mathscr{A} }_2  = \{(x,y):  x,y \in\{1,\ldots,5\} \} \times \{1,\ldots,300\}.$}
	
\end{itemize}

\begin{figure} [h!]
	\centering
	\includegraphics[scale=0.35]{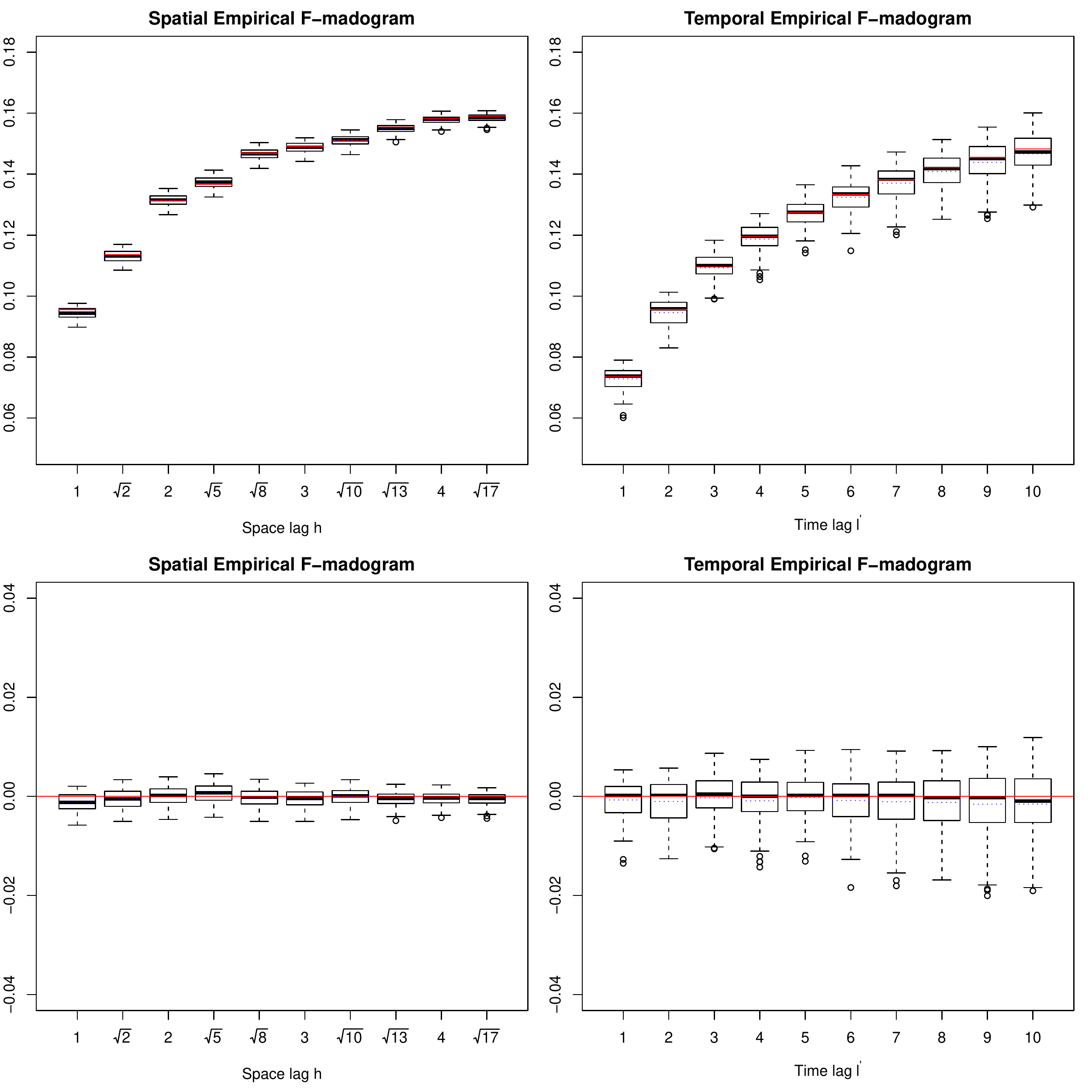}
	
	\caption{Scheme 1: (Top row) boxplots of purely spatial/temporal empirical $F$-madograms estimates at lags $(h,l') \in \mathcal{H} \times \mathcal{K}$ for 100 simulated BR processes (\ref{BR spectral fun}) with FBM spatio-temporal semivariogram (\ref{FBM semi}). The middle blue dotted/red solid lines show the overall mean of the estimates/true values. (Bottom row) boxplots of the corresponding estimation errors.  } 
	
	\label{Empirical mado}
	
\end{figure}

\begin{figure} [!h]
	\centering
	\includegraphics[scale=0.50]{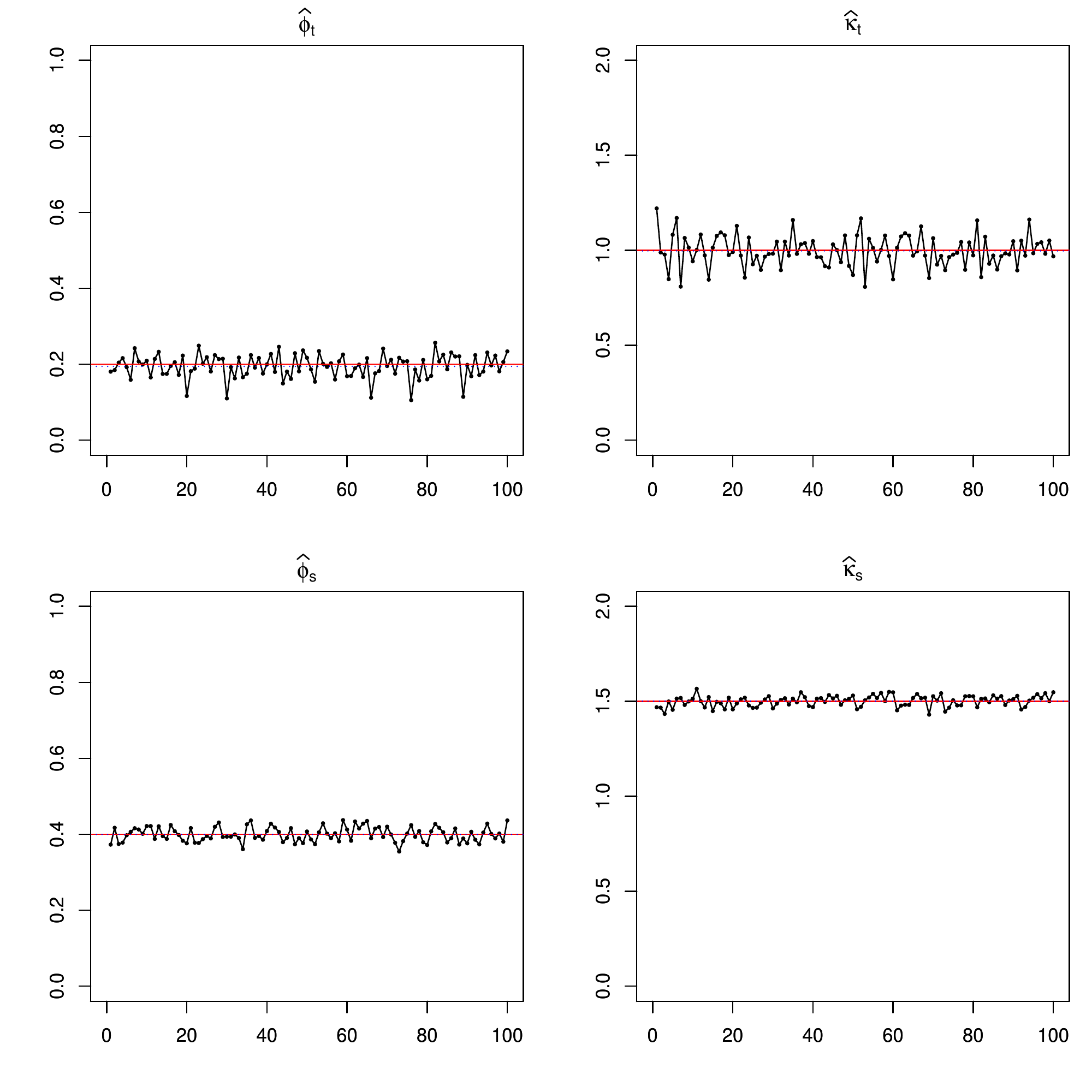}
	
	\caption{Scheme 1: Semiparametric estimates of $\boldsymbol{\widehat{\psi}} = \{  \widehat{\phi}_s, \widehat{\kappa}_s, \widehat{\phi}_t,\widehat{\kappa}_t\}$ for 100 simulated BR processes defined by (\ref{BR spectral fun}) with FBM spatio-temporal semivariogram (\ref{FBM semi}). The middle blue dotted/red solid lines show the overall mean of the estimates/true values.} 
	
	\label{Performance1}
	
\end{figure}

We assess the quality of the fit between the theoretical values of spatial/temporal $F$-madograms and their estimates. Figure~\ref{Empirical mado} compares empirical estimates of purely spatial/temporal $F$-madograms with their asymptotic counterparts. Overall, both the purely spatial/temporal empirical versions are consistent, with a relatively higher variability for the temporal estimates. This is probably due to the fairly low number of time instants (300) used for the estimation of the purely temporal parameters compared to the number of spatial locations (2500) used for the estimation of the purely spatial parameters.
Next, we present results for the semiparametric estimation with Scheme1. Figure~\ref{Performance1} displays the resulting estimates of the purely spatial parameters $(\phi_s,\alpha_s)$ and the purely temporal parameters $(\phi_t,\alpha_t)$. Generally, the estimation procedure appears to work well. Moreover, we observe that the estimation of the purely spatial parameters is more accurate (the RMSE and MAE are lower), see Table \ref{BR}. Again this probably stems from the large number of spatial locations used in the estimation which is ($\approx 8.3$) times higher than the time points.

As the last step in this study, we compare the statistical efficiency of our method and the one proposed in \cite{buhl2016semiparametric}. Table \ref{BR} reports the performance metrics for both methods. Although in that study, the authors used a larger grid size $(n=70)$ to estimate the purely spatial parameters, clearly, the $F$-madogram semiparametric estimation outperforms their approach which based on the extremogram as an inferential tool (their semiparametric estimates show a larger bias than ours; the RMSE and MAE are higher). This is probably due to the fact that the estimates obtained in \cite{buhl2016semiparametric} are sensitive to the choice of the threshold used for computing (possibly bias corrected) empirical estimates of the extremogram.

\subsubsection{Estimation using Scheme 2} \label{Sec::scheme1.2}
\begin{figure} [h!]
	\centering
	\includegraphics[scale=0.34]{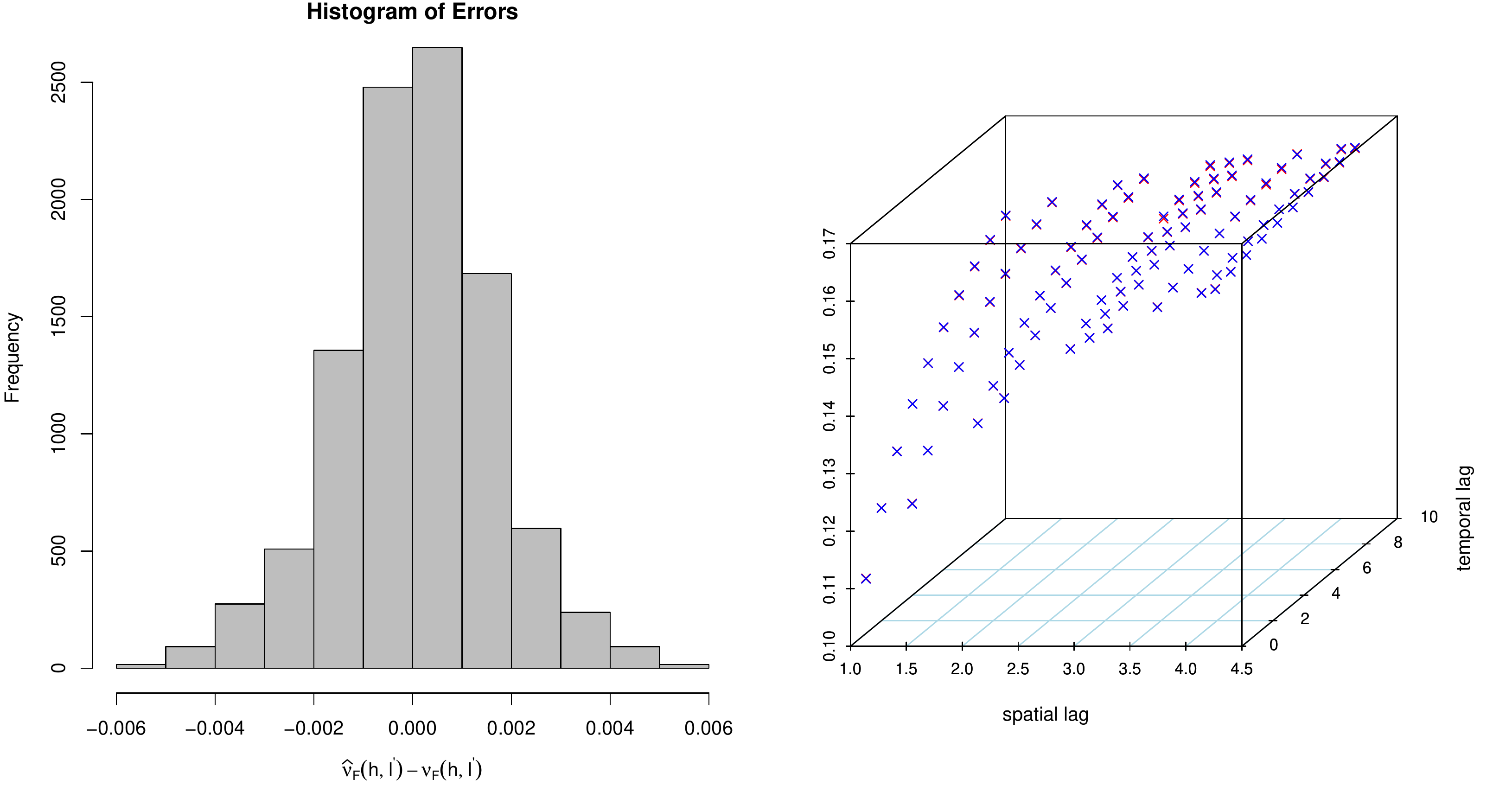}
	
	\caption{Scheme 2: Diagnostic plots of the empirical spatio-temporal $F$-madogram estimates for 100 simulated BR processes defined by (\ref{BR spectral fun}) with FBM spatio-temporal semivariogram (\ref{FBM semi}). Histogram of the errors, $\widehat{\nu}_F(h,l')-\nu_F(h,l')$, $(h,l') \in \mathcal{H} \times \mathcal{K}$ (left panel). Blue/red cross symbols show the overall mean of the empirical spatio-temporal $F$-madogram estimates/model-based counterparts (right panel). } 
	
	\label{3Dspatiomado}
	
\end{figure}

Based on Scheme 2, we estimate the parameters of the space-time max-stable BR process with a similar simulation setting which is previously described in Section~\ref{Sec::setup1}. We consider the space-time observation area where the spatial locations consisted of a $20 \times 20$ grid and equidistantly time points, $\{1,\ldots,200\}$. Figure~\ref{3Dspatiomado} compares the empirical spatio-temporal $F$-madogram estimates $\widehat{\nu}_F(h,l')$ with their model-based counterparts $\nu_F(h,l')$ over the spatio-temporal lags $(h,l') \in \mathcal{H} \times \mathcal{K}$. There is a good agreement overall. These diagnostic plots provide a satisfactory representation of the empirical spatio-temporal $F$-madogram estimates. Generally, the results lend support to the agreement between the empirical spatio-temporal $F$-madogram estimates and model-based counterparts, especially once sampling variability is taken into account.

\begin{figure} [h!]
	\centering
	\includegraphics[scale=0.45]{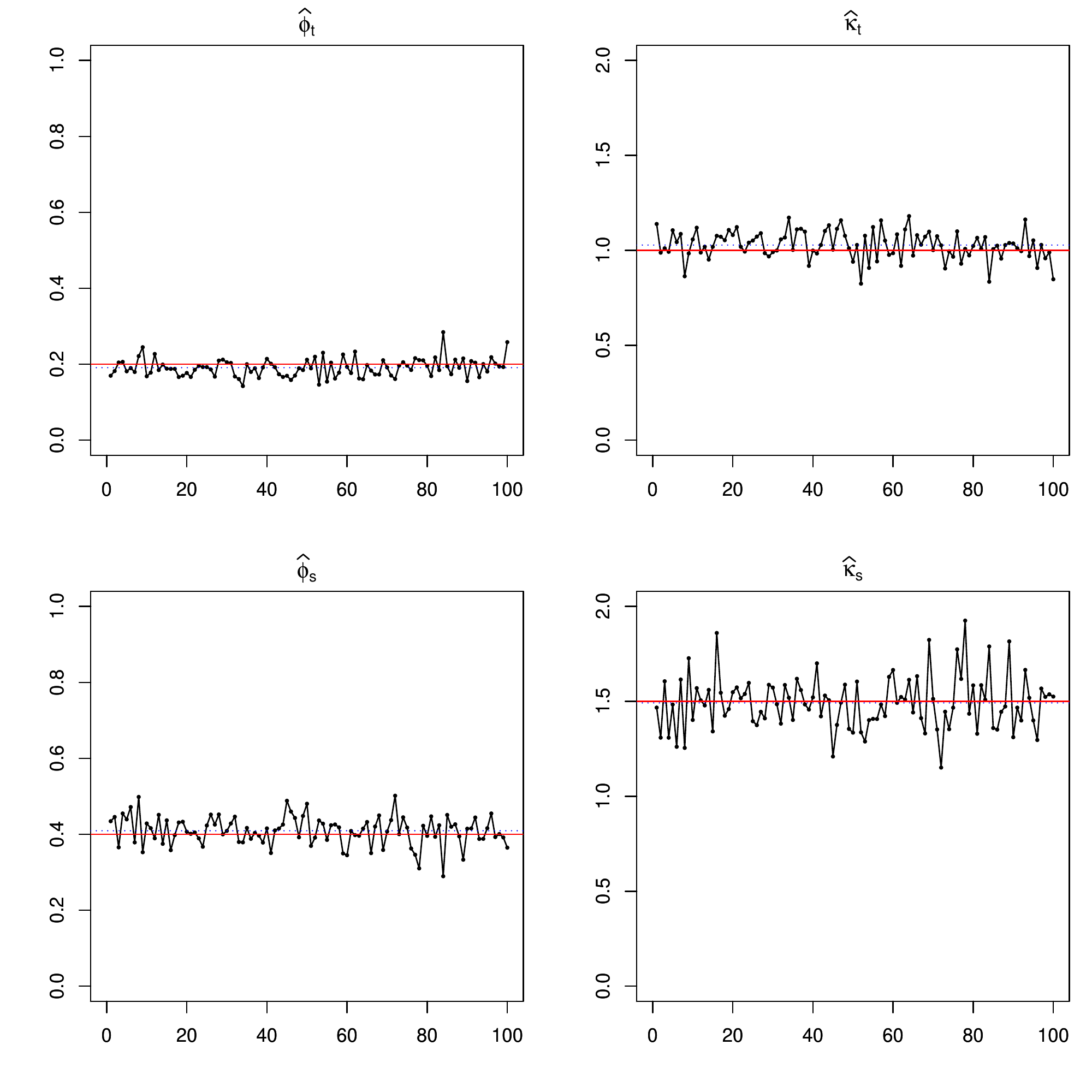}
	
	\caption{Scheme 2: Semiparametric estimates of $\boldsymbol{\widehat{\psi}} = \{ \widehat{\phi}_s, \widehat{\kappa}_s,\widehat{\phi}_t,\widehat{\kappa}_t\}$ for 100 simulated BR processes defined by (\ref{BR spectral fun}) with FBM spatio-temporal semivariogram (\ref{FBM semi}). The middle blue dotted/red solid lines show overall mean of the estimates/true values. } 
	
	\label{BRscheme2}
	
\end{figure}
Figure~\ref{BRscheme2} shows the estimation performance of the estimated parameters. Overall, the parameters are well estimated. Moreover, we observe that the estimation of the scale parameters $\{\phi_s$, $\phi_t\}$ is more accurate than the smoothness parameters $\{\kappa_s, \kappa_t\}$ (the RMSE and MAE are lower), see Table \ref{BR}.

\begin{figure} [htp!]
	\centering
	\includegraphics[scale=0.5]{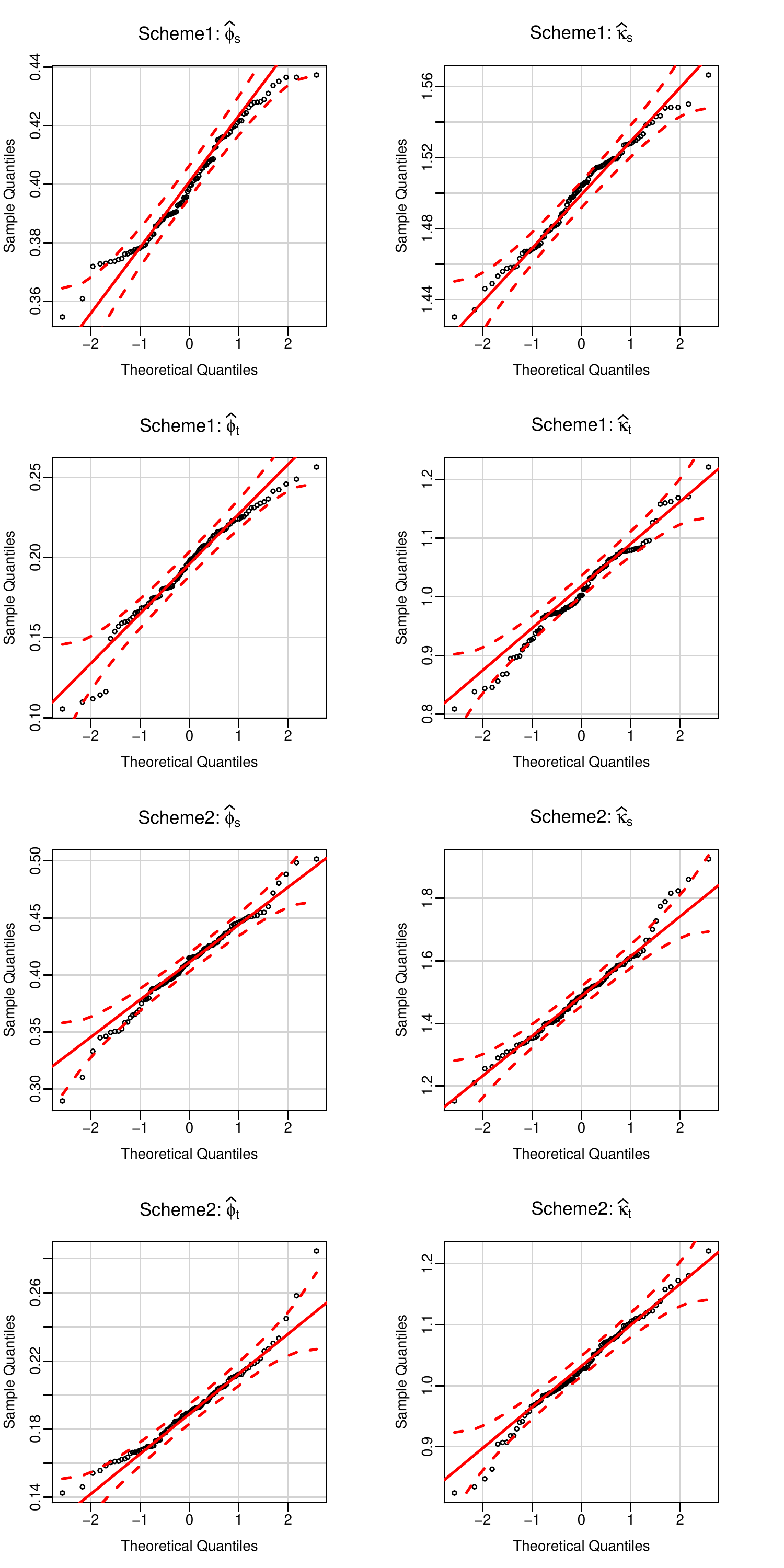}
	
	\caption{QQ-plots of the estimates resulting from both estimation schemes for 100 simulated BR processes defined by (\ref{BR spectral fun}) with the FBM spatio-temporal semivariogram (\ref{FBM semi}) against the normal distribution. Scheme 1: purely spatial parameters (top row) and purely temporal parameters (second row). Scheme 2: purely spatial parameters (third row) and purely temporal parameters (bottom row). Dashed red lines correspond to 95\% confidence intervals. } 
	
	\label{QQBR}
	
\end{figure}

To sum up, for both schemes, Table \ref{BR} reports the mean estimate, RMSE, and MAE of the estimated parameters 
$\boldsymbol{\widehat{\psi}} = \{ \widehat{\phi}_s, \widehat{\kappa}_s,\widehat{\phi}_t,\widehat{\kappa}_t\}$. Let us remark that the comparison between the resulting parameter estimates from the two estimation schemes is not completely straightforward because we consider non-unified space-time observation areas due to the above-mentioned computational reasons. However, with the above sampling schemes, we observe that the estimation of the purely spatial parameters is more accurate when using Scheme 1 (the RMSE and MAE are lower). On the other hand, we notice a slight outperformance for Scheme 2 in estimating purely temporal parameters. Finally, the QQ-plots against a normal distribution in Figure~\ref{QQBR} provide an indication for asymptotic normality of the resulting estimates.
\begin{table*}[b!]
	\centering
	\scalebox{0.72}{
		\begin{tabular}{@{}l|lll | llll| llll@{}}\hline
			& \multicolumn{3}{c|}{Scheme 1}&& \multicolumn{3}{c|}{Scheme 1, \cite{buhl2016semiparametric}}& & \multicolumn{3}{c}{Scheme 2} \\
			
			\cline{2-4}
			\cline{5-8}
			\cline{9-12}
			True& Mean estimate  & RMSE & MAE && Mean estimate & RMSE& MAE &&Mean estimate & RMSE& MAE  \\ \hline

			Purely Spatial & & & & & & && &   &&\\
			${\phi}_s=0.4$ & 0.3998&0.0191& 0.0162&& 0.4033 &0.0678 & 0.0559  && 0.4093&0.0389& 0.0307\\
			${\kappa}_s=1.5$ & 1.5019&0.0289& 0.0243&&1.4984 & 0.0521& 0.0400  &&    1.4921&0.1399&  0.1083\\
			
			\hline

			Purely temporal & & & & & & && &   &&\\
			${\phi}_t=0.2$& 0.1944 & 0.0314 & 0.0246 &&0.2249 &0.0649 & 0.0526  &&0.1909& 0.0251&0.0201\\
			${\kappa}_t=1$ & 0.9969& 0.0831&0.0657&& 0.9563&0.0939 & 0.0767  &&1.0278& 0.0785&0.0619\\
			\hline
		\end{tabular}
	}
	\caption{Performance of the estimation for 100 simulated BR processes defined by (\ref{BR spectral fun}) with FBM spatio-temporal semivariogram (\ref{FBM semi}). The mean estimate, RMSE, and MAE of the estimated parameters.}
	\label{BR}
\end{table*}


\subsection{Simulation study 2: Fitting spectrally separable space-time max-stable Smith process} \label{Sec::sim2}

\subsubsection{Setup for a simulation study} \label{Sec::setup2}
We simulate data from the spatio-temporal Smith process considered in Example~\ref{example2}, with parameter vector $\boldsymbol{\psi}=(1,0, 1, 1,1,0.7)^{t}$. As a reasonable compromise between accuracy and computation time, the locations are assumed to lie on a regular 2D grid of size $n=20$. The time points are equidistant, given by the set $\{1,\ldots,200\}$. The simulations have been carried out using R SpatialExtremes package with {\em rmaxstab} function, see \cite{ribatet2011spatialextremes}. The spatial lags set $\mathcal{H}$ and temporal lags set $\mathcal{K}$ are fixed as before, recall Section~\ref{Sec::sim1}. Equal weights are assumed. We repeat this experiment 100 times.

\subsubsection{Results for the two estimation  schemes} \label{Sec::scheme2.1}

\begin{figure} [h!]
	\centering
	\includegraphics[scale=0.4]{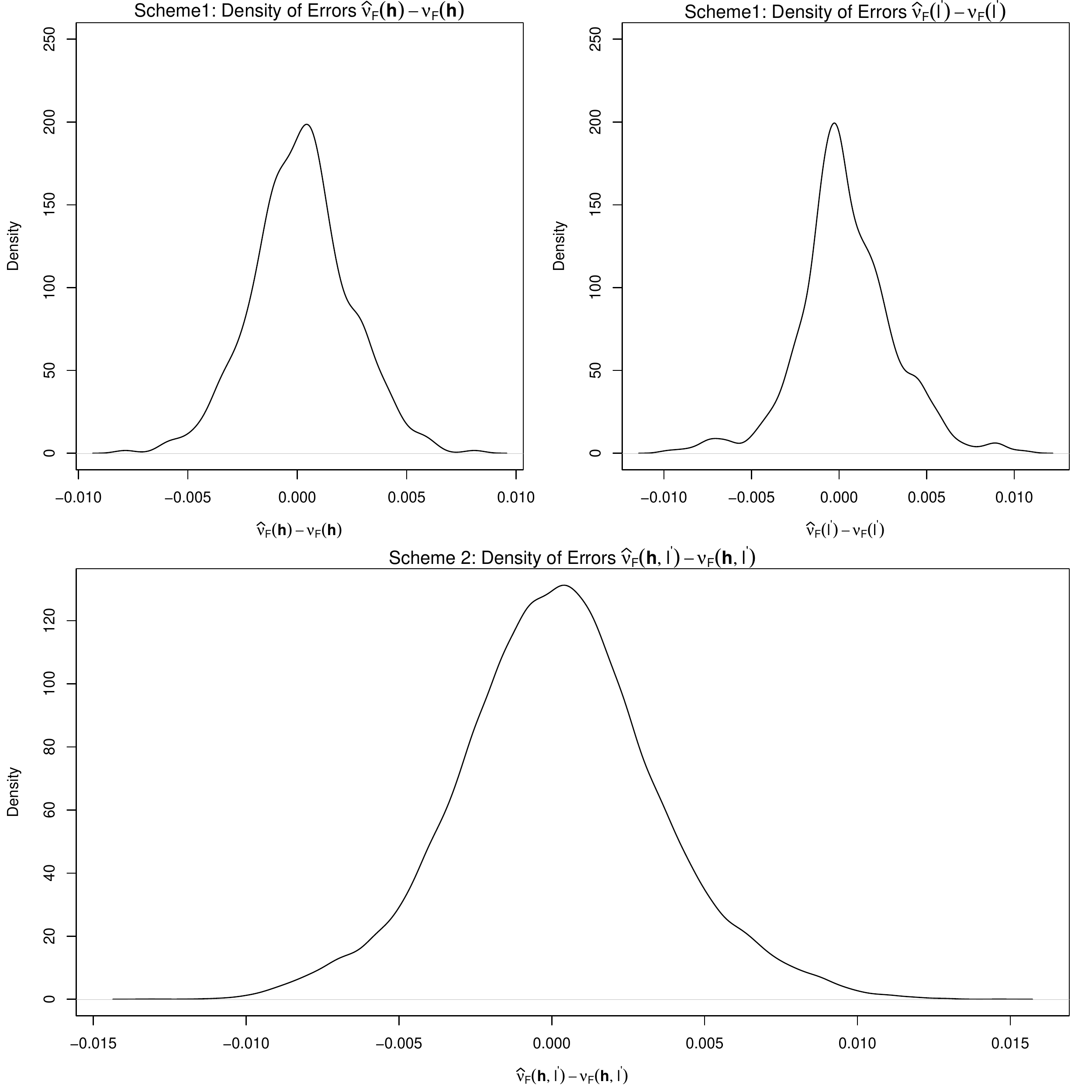}
	
	\caption{Density of the errors between the empirical versions of the $F$-madogram estimates and their model-based counterparts for 100 simulated spectrally separable space-time max-stable Smith processes with parameter $\boldsymbol{\psi}=(1,0, 1, 1,1,0.7)^{t}$. Scheme 1 (Top row): ${\widehat{\nu}}_{F} (\boldsymbol{h}) -{\nu}_{F} (\boldsymbol{h})$, $\lVert\boldsymbol{h}\rVert \in \mathcal{H}$ (left panel) ${\widehat{\nu}}_{F} (l') - {\nu}_{F} (l')$, $l' \in \mathcal{K}$ (right panel). Scheme 2 (Bottom row):  $\widehat{\nu}_F(\H,l') - {\nu}_F(\H,l'),$ at spatio-temporal lags $(\HH,l') \in \mathcal{H} \times \mathcal{K}$. } 
	
	\label{Density}
	
\end{figure}

The top row of Figure~\ref{Density} displays the density of the errors between the empirical estimates of the purely spatial/temporal $F$-madograms and their model-based counterparts, whereas the bottom row displays the density of the errors between empirical spatio-temporal $F$-madogram estimates and model-based counterparts. Generally, all of the empirical versions are congruous with their asymptotic counterparts. Clearly, the density of the errors is close to a centered Gaussian distribution. 

\begin{figure} [h!]
	\centering
	\includegraphics[scale=0.35]{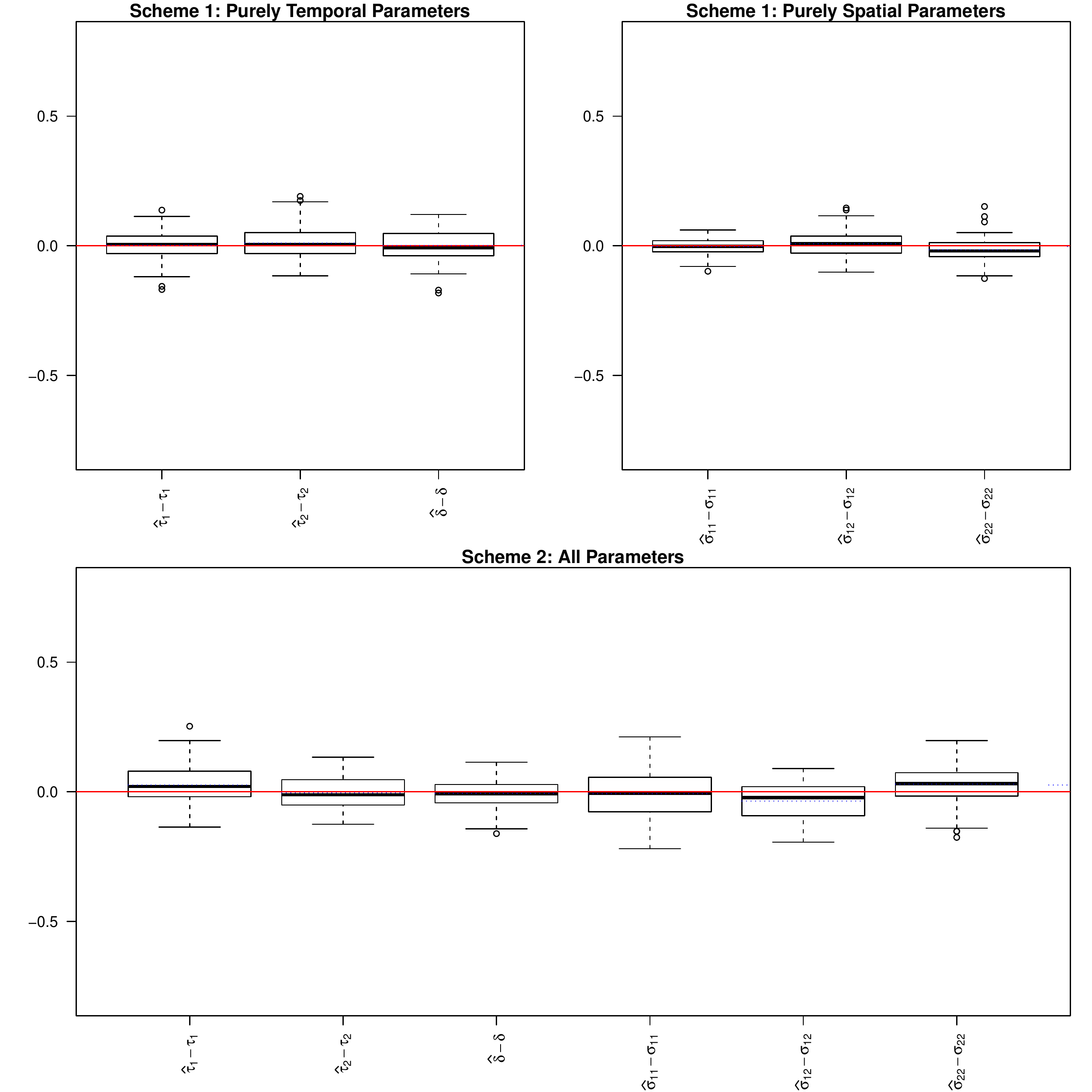}
	
	\caption{Boxplots of the errors $\boldsymbol{\widehat{\psi}}-\boldsymbol{{\psi}}$ resulting from both estimation schemes for 100 simulated spectrally separable space-time max-stable Smith processes with parameter $\boldsymbol{\psi}=(1,0, 1, 1,1,0.7)^{t}$. Scheme 1 (Top row): purely spatial parameters (left panel) and purely temporal parameters (right panel). Scheme 2 (Bottom row): all parameters. The middle blue dotted/red solid lines show the overall mean of errors estimates/zero value.} 
	
	\label{Smithall}
	
\end{figure}

Figure~\ref{Smithall} displays boxplots the errors of the resulting estimates from both schemes: ($\boldsymbol{\widehat{\psi}}-\boldsymbol{{\psi}}$). The top row displays the estimation errors of purely spatial parameters $({\sigma}_{11},{\sigma}_{12},{\sigma}_{22})$ and purely temporal parameters $({\tau}_1,{\tau}_2,{\delta})$ resulting from Scheme 1, whereas the bottom row displays the estimation errors resulting form Scheme 2. Overall, the inference procedures perform well. Altogether, we observe that the estimates are close to the true values.

To sum up, for both schemes, Table \ref{ST Smith} reports the mean estimate, RMSE, and MAE of the estimated parameters 
$\boldsymbol{\widehat{\psi}} = \{\widehat{\sigma}_{11},\widehat{\sigma}_{12},\widehat{\sigma}_{22},\widehat{\tau}_1,\widehat{\tau}_2,\widehat{\delta} \}$. Contrary to Scheme 2, we observe that the estimation of purely spatial parameters $\boldsymbol{\Sigma}$ is more accurate than the estimation of purely temporal parameters ($\boldsymbol{\tau}$ and $\delta$) when using Scheme 1 (RMSE and MAE are lower). This probably can be justified by the fact that in Scheme 1 the number of spatial locations used is higher than time moments. Additionally, there is probably an impact of the estimated covariance matrix $\widehat{\boldsymbol{\Sigma}}$ on the estimation efficiency of the purely temporal parameters, whereas, the purely temporal parameters are estimated independently of purely spatial parameters when using Scheme 2.  Moreover, we notice that the estimation of purely spatial parameters is less accurate when using Scheme 2 (RMSE and MAE are higher). This is probably owing to the fact that in Scheme 2 the number of pairs used is higher than in Scheme 1, leading more variability. Whereas, both schemes seem to have the same performance order in estimating purely temporal parameters.

\begin{figure} [h!]
	\centering
	\includegraphics[scale=0.5]{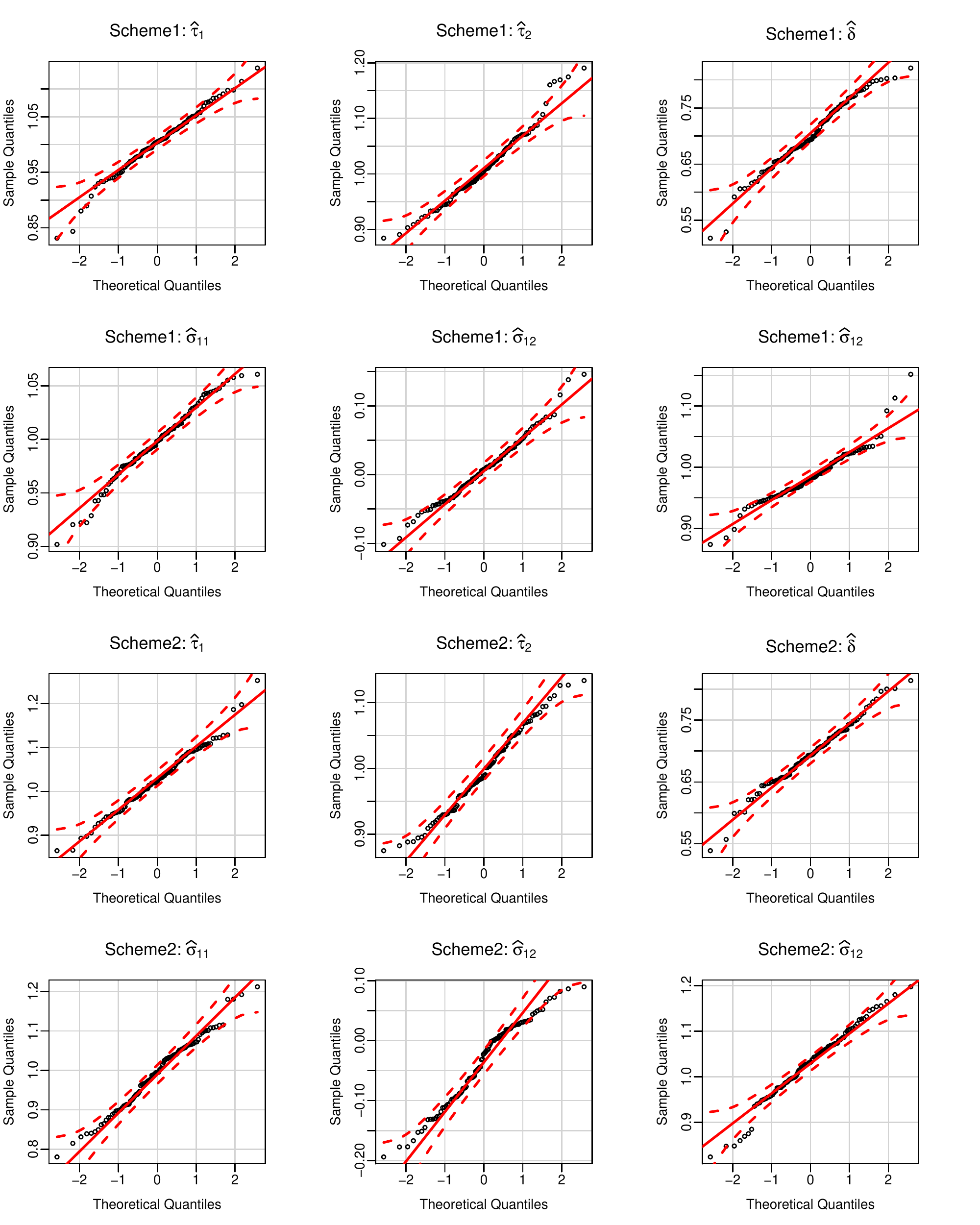}
	
	\caption{QQ-plots of the estimates from both estimation schemes for 100 simulated spectrally separable space-time max-stable Smith processes with parameter $\boldsymbol{\psi}=(1,0, 1, 1,1, 0.7)^{t}$ against the normal distribution. Scheme 1: purely spatial parameters (top row) and purely temporal parameters (second row). Scheme 2: purely spatial parameters (third row) and purely temporal parameters (bottom row). Dashed red lines correspond to 95\% confidence intervals. } 
	
	\label{qqplot}
	
\end{figure}

We also show QQ-plots against a normal distribution for all parameters in Figure~\ref{qqplot}. For both schemes, it seems that the semiparametric estimates are approximately normally distributed.

\begin{table*}[htp!]
	\centering
	\scalebox{0.95}{
		\begin{tabular}{@{}l|lll | llll@{}}\hline
			& \multicolumn{3}{c|}{Scheme 1} & & \multicolumn{3}{c}{Scheme 2} \\
			
			\cline{2-4}
			\cline{5-8}
			True& Mean estimate  & RMSE & MAE && Mean estimate & RMSE& MAE  \\ \hline 
			
			Purely Spatial & & & & & & &\\
			$\sigma_{11}=1$&0.9973  &         0.0331       &     0.0259  &&   0.9929      &  0.0888  & 0.0727  \\
			${\sigma}_{12} =0$&  0.0081 &   0.0470       &    0.0369   &&  $-$0.0357 & 0.0770 &0.0609   \\
			${\sigma}_{22}=1 $ &0.9848   &  0.0440      &    0.0346   &&   1.0295&  0.0805 & 0.0647\\

			\hline
			
			Purely temporal & & & & & & &\\
			${\tau}_1=1$ & 1.0021               & 0.0549  &    0.0426              &&        1.0261               &  0.0747    & 0.0591 \\
			${\tau}_2=1$ &  1.0107              &   0.0646    &0.0505                 &&    0.9962               & 0.0620  & 0.0516    \\
			$\delta=0.7$ &  0.7012               &     0.0595    &0.0482       &&     0.6939                    &   0.0510         &    0.0400   \\
			
			\hline
		\end{tabular}
	}
	
	\caption{Performance of the estimation for 100 simulated spectrally separable space-time max-stable Smith processes considered in Example~\ref{example2}, with parameter $\boldsymbol{\psi}=(1,0, 1, 1,1,0.7)^{t}$. The mean estimate, RMSE, and MAE of the estimated parameters.}
	\label{ST Smith}
\end{table*}

Finally, let us remark that a simulation study has been carried out in \cite{embrechts2016space}, where only the spectrally separable spatio-temporal Smith process has been fitted. Irregularly spaced locations have been considered. Two estimation schemes based on pairwise likelihood have been adopted (a two-step approach and a one-step approach). The obtained results have shown that, the estimation of purely spatial parameters is more accurate with a two-step approach.


\subsection{Simulation study 3: Fitting spectrally separable STMS Schlather process}\label{app4}

\begin{figure} [h!]
	\centering
	\includegraphics[scale=0.35]{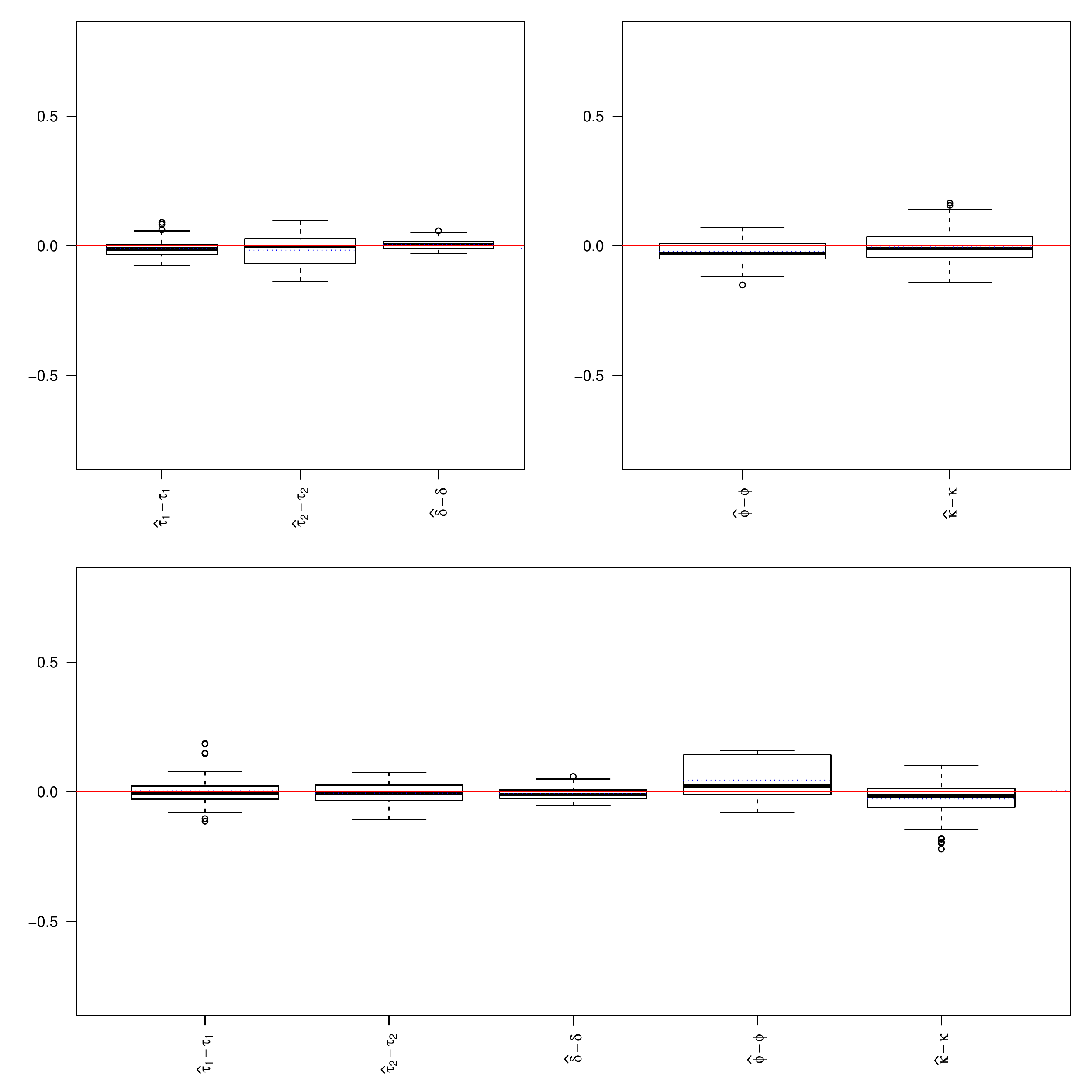}
	
	\caption{Boxplots of errors $\boldsymbol{\widehat{\psi}}-\boldsymbol{{\psi}}$ from both estimation schemes for 100 simulated spectrally separable STMS Schlather processes with parameter $\boldsymbol{\psi}=(2,1.5, 1,0,0.3)^{t}$. Scheme 1 (Top row): purely spatial parameters (left panel) and purely temporal parameters (right panel). Scheme 2 (Bottom row): all  parameters. The middle blue dotted/red solid lines show the overall mean of errors estimates/zero value. } 
	
	\label{Sch}
	
\end{figure}

Finally, we perform a third simulation study to fit spectrally separable space-time max-stable Schlather process. The innovation process $H$ is derived from independent replications of a spatial Schlather process with correlation function of powered exponential type defined, for all $\HH \geq0$, by $\rho(\H)=\exp [-  ({\HH} / \phi)^{\kappa}]$, $\phi>0$ and $0<\kappa<2$, where $\phi$ and $\kappa$ denote, respectively, the range and the smoothing parameters. We denote by $\boldsymbol{\psi}=(\phi,\kappa,  \tau_1,\tau_2,\delta)^{t}$ the vector gathering the model parameters. We take $\phi=3$, $\kappa=3/2$, $\boldsymbol{\tau}=(1,0)^{t}$ and $\delta=0.3$ . As previously, we consider the same simulation setup used in Section~\ref{Sec::setup2}. The results are summarized in Figure~\ref{Sch} and Table~\ref{Schlather}. Generally, we obtain equally satisfying results.


\begin{table*}[htp!]
	\centering
	\scalebox{0.85}{
		\begin{tabular}{@{}l|lll | llll@{}}\hline
			& \multicolumn{3}{c|}{Scheme 1} & & \multicolumn{3}{c}{Scheme 2} \\
			
			\cline{2-4}
			\cline{5-8}
			True& Mean estimate  & RMSE & MAE && Mean estimate & RMSE& MAE  \\ \hline 
			Purely Spatial & & & & & & &\\
			$\phi=2$              & 1.9841  & 0.0368     &     0.0309       &&    2.0357     & 0.0812    & 0.0599  \\
			${\kappa}=1.5$    & 1.4967  & 0.0407       &   0.0327         &&     1.4814   & 0.0771   &   0.0557\\

			\hline
			
			Purely temporal & & & & & & &\\
			${\tau}_1=1$    &   0.9852       &    0.0442       &   0.0346     &&  1.0036          & 0.0556    & 0.0393 \\
			${\tau}_2=0$ &     $-$0.0177  &  0.0636    &        0.0512    &&    $-$0.0053  &  0.0427    &  0.0353 \\
			$\delta=0.3$        &     0.3031      &     0.0473    &     0.0383    &&  0.2913        &  0.0393    & 0.0318 \\
			
			\hline
		\end{tabular}
	}
	
	\caption{Performance of the estimation for 100 simulated spectrally separable STMS Schlather processes, with parameter $\boldsymbol{\psi}=(2,1.5, 1, 0,0.3)^{t}$. The mean estimate, RMSE, and MAE of the estimated parameters.}
	\label{Schlather}
\end{table*}


\section{Real data analysis} \label{sec:real}
In this section, we aim to quantify the extremal behavior of radar rainfall data in a region in the State of Florida. Our approach is to fit the data by different space-time max-stable classes based on a space-time block maxima design using the proposed semiparametric estimation procedure. 

\subsection{Description of the dataset}
\begin{figure} [h!]
	\centering
	\includegraphics[scale=0.25]{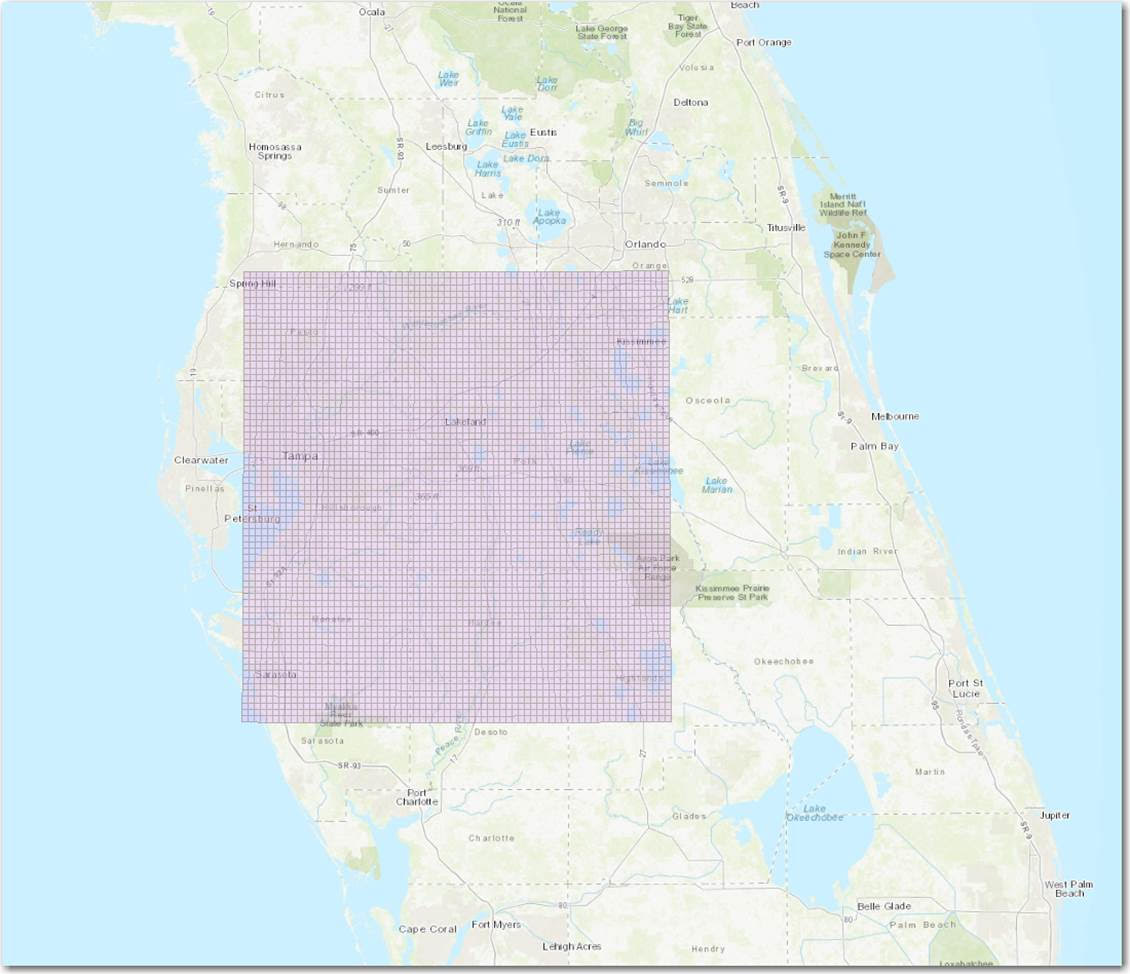}
	
	\caption{Radar rainfall observation area in the State of Florida. Source: Southwest Florida Water Management District (SWFWMD).} 
	
	\label{Grid}
	
\end{figure}

The dataset analyzed in this section is composed of radar rainfall values (in inches) measured on a square of 140 $\times$ 140 km region containing 4900 grid locations in the State of Florida. The database consists of radar hourly rainfall values measured on a regular grid with squared cells of size 2 km covering a region of 70 $\times$ 70 cells in the State of Florida. A map of the study area is shown in Figure~\ref{Grid}. We only consider the wet season (June-September) over the years 2007-2012. The data were collected by the Southwest Florida Water Management District (SWFWMD) and freely available on \url{ftp://ftp.swfwmd.state.fl.us/pub/radar_rainfall}. Moreover, the dataset is available in the Supplementary Material:  		\url{http://math.univ-lyon1.fr/homes-www/abuawwad/Florida_RadarRainfall/}. 

\subsection{Data fitting}
\begin{figure} [h!]
	\centering
	\includegraphics[scale=0.48]{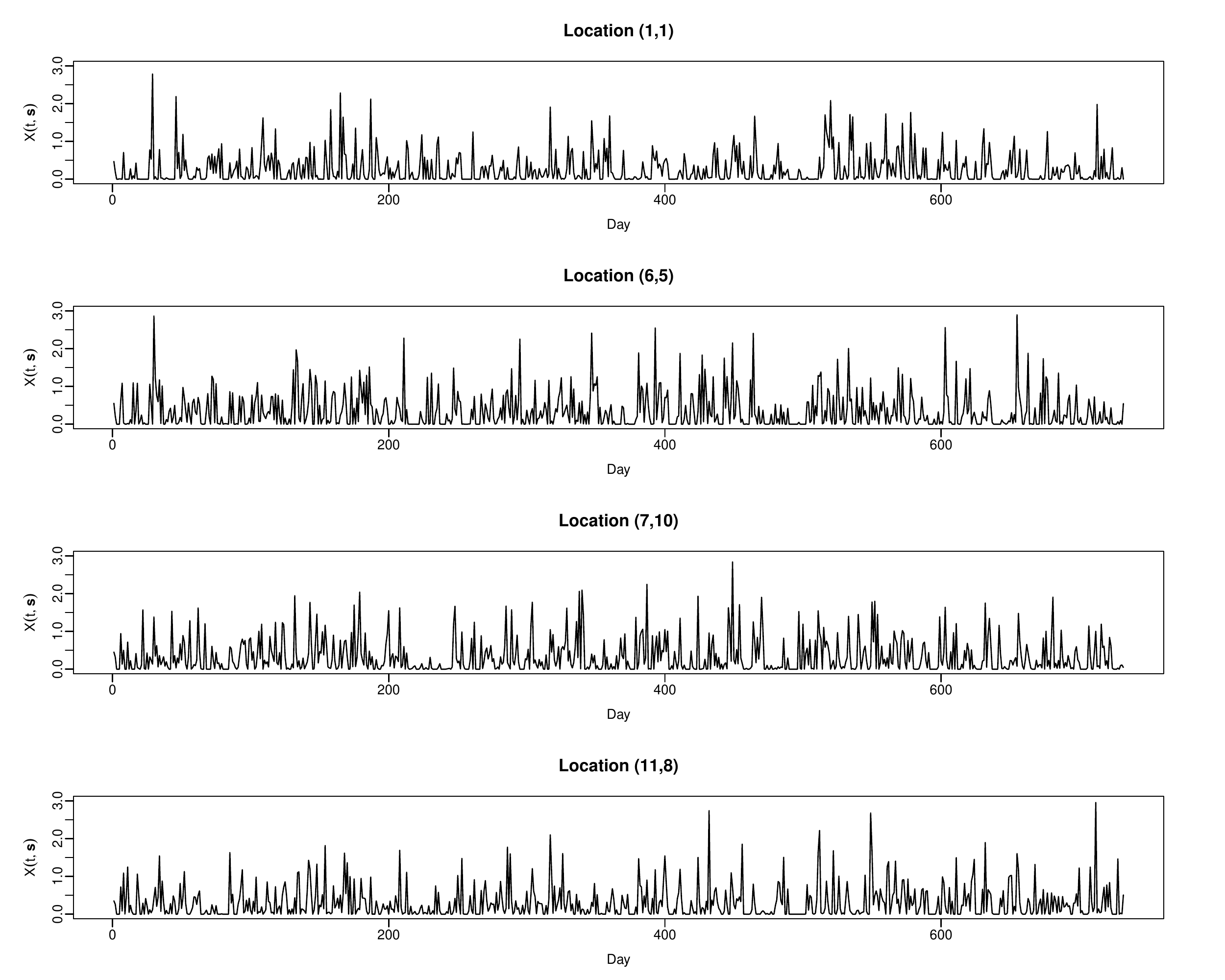}
	
	\caption{Plots of daily maximal rainfall in inches for four grid locations with simplified coordinates: (1,1), (6,5), (7,10) and (11,8).} 
	
	\label{timeseries}
	
\end{figure}

We perform a block maxima design in space and time as follows: we take block maxima over 24 consecutive hours and over 10 km $\times$ 10 km areas (the daily maxima over 25 grid locations), resulting in $14 \times 14$ grid in space for all $6 \times 122$ days of the wet seasons. So, this gives a time series of dimension $14 \times 14$ and of length 732. For the sake of notational simplicity, we denote the set of resulting grid locations by $\mathbb{S} = \left\{(x,y):  x,y \in\{1,\ldots,14\} \right\}$ and the spacetime
realizations by $\left\{X(\S,t),\boldsymbol{s} \in \mathbb{S} , \ t\in \{t_1,\ldots,t_{732}\} \right\}$. This setup has been also considered in \cite{buhl2016semiparametric,davis2013statistical} for analyzing radar rainfall measurements in a region in the State of Florida over the years 1999-2004, where only space-time max-stable BR process has been fitted to the data by a semiparametric approach in \cite{buhl2016semiparametric} and a pairwise likelihood approach in \cite{davis2013statistical}. Let us remark that both regions here and in the above-mentioned two studies are located in the central portion of Florida District, which would probably have the best square area of coverage. Having larger grid size will lead to some cells missing in the southwestern \enquote*{corner} due to the coastline. Figure~\ref{timeseries} shows the obtained time series for daily maxima observations at four grid locations.
	
	
	
According to this sampling scheme of the process $X$, there are $196 \times 731= 143276$ spatio-temporal pairs of points at distance $(0,1)$, that is,
$$ \left\{ (\S_1,t_2), (\S_1,t_1) \right\}, \left\{ (\S_2,t_2), (\S_2,t_1) \right\}, \ldots, \left\{ (\S_{196},t_2), (\S_{196},t_1) \right\}  $$
$$\vdots $$
$$ \left\{ (\S_1,t_{732}), (\S_1,t_{731}) \right\}, \left\{ (\S_2,t_{732}), (\S_2,t_{731}) \right\}, \ldots, \left\{ (\S_{196},t_{732}), (\S_{196},t_{731}) \right\}.$$
Analogously, there are $196 \times 730= 143080$ spatio-temporal pairs of points at distance $(0,2)$, and so forth. Generally, for a set of spatio-temporal data measured in the time moments $t_1, \ldots, t_T$, on a regular $n \times n$ spatial grid, we have $n^2(t_T-l')$ spatio-temporal pairs of points at distance $(0,l')$. Computing the $F$-madogram values corresponding to the above spatio-temporal distances, we obtain the purely temporal empirical $F$-madogram. It is also easy
to check that there are $364 \times 732 = 266448$ spatio-temporal pairs of points at distance $(1,0)$, $336 \times 732 = 245952$ at distance $(2,0)$, and so forth, see Table~\ref{Lags}. Computing the $F$-madogram values for the spatio-temporal distances $(\H,0)$, we obtain the purely spatial empirical $F$-madogram.
\begin{table} [htp!]
	
	\centering
	\setlength\tabcolsep{1pt} 
	\scalebox{0.82}{
		\begin{tabular}{|c | c | }
			\hline
			Distance $(h, 0)$& Number of spatio-temporal pairs of points  \\
			\hline	\hline
			$(1,0)$ &  $2n(n-1) \times t_T$\\ 
			$\left(\sqrt{2} ,0\right)$ & $2(n-1)^2 \times t_T$ \\ 
			$(2,0)$ & $2n(n-2) \times t_T$\\ 
			$\left(\sqrt{5} ,0\right)$ & $4 (n-1) (n-2) \times t_T$\\ 
			$\left(\sqrt{8},0 \right)$ & $2(n-2)^2 \times t_T$\\ 
			$(3,0)$ & $2n(n-3) \times t_T$\\ 
			$\left(\sqrt{10},0 \right)$ &$4 (n-1) (n-3) \times t_T$ \\ 
			
			$\left(\sqrt{13} ,0\right)$ &$4 (n-2) (n-3) \times t_T$ \\ 
			$(4,0)$ & $2n(n-4) \times t_T$\\ 
			$\left(\sqrt{17},0 \right)$ & $4 (n-1) (n-4) \times t_T$\\ 
			\hline
			
		\end{tabular}
	}
	\caption{Number of spatio-temporal points at distance $(0, h)$ for a set of spatio-temporal data measured in the time moments $t_1, \ldots, t_T$, on a regular $n \times n$ spatial grid.}
	
	\label{Lags}
\end{table}

\begin{figure} [htp!]
	\centering
	\includegraphics[scale=0.35]{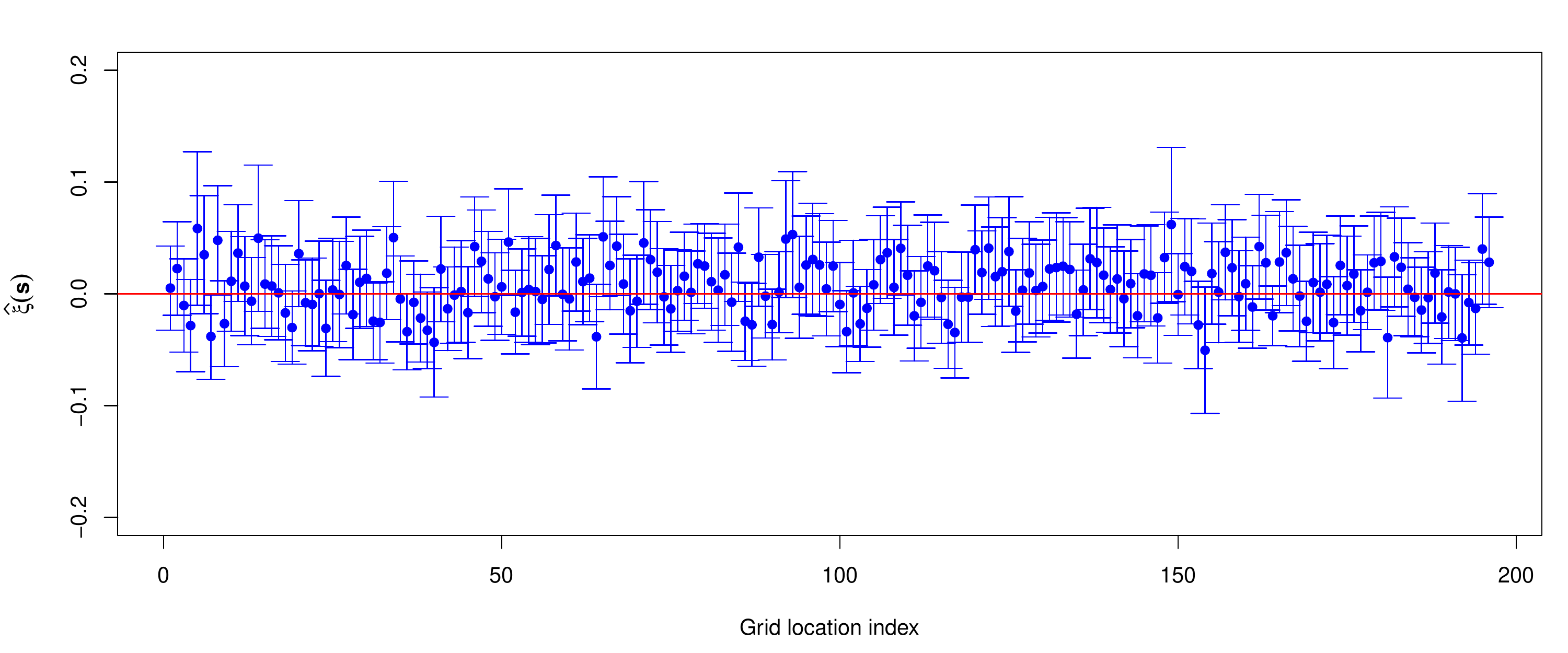}
	
	\caption{Estimated GEV shape parameter $\widehat{\xi}{(\S)}$ at all grid locations with 95\% confidence intervals. } 
	
	\label{RADARFIT}
	
\end{figure}

\begin{figure} [h!]
	\centering
	\includegraphics[scale=0.4]{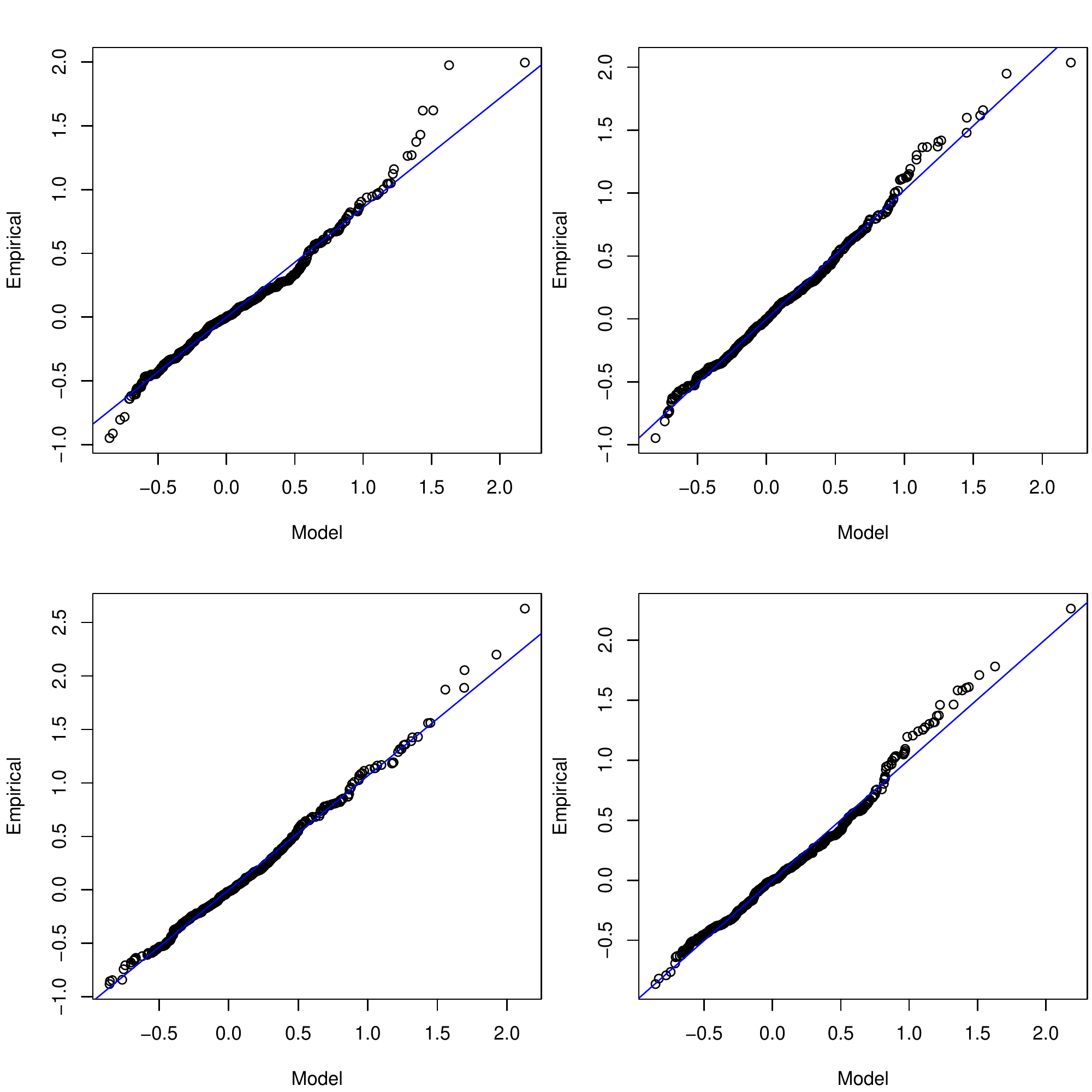}
	
	\caption{QQ-plots of deseasonalized rain series versus the fitted Gumbel distribution (GEV with $\widehat{\mu}(\S)$, $\widehat{\sigma}(\S)$ and 0) on the basis of the time series corresponding to the four grid locations shown in Figure~\ref{timeseries}.} 
	
	\label{qqplotGumbel}
	
\end{figure}

Since we are interested in modeling the joint occurrence of extremes over a region, then the dependence structure
of a multivariate variable has to be explicitly stated. The usual modeling strategy consists of two steps: firstly, estimating the marginal distribution. Secondly, characterizing the dependence via a model issued by the multivariate
extreme value theory, see e.g., \cite{beirlant2006statistics,padoan2010likelihood}. For marginal modeling, we explain the procedure as follows: 
\begin{enumerate}[label=(\roman*)]
	\item We transform the data to stationarity by removing possible seasonal effects using a simple moving average with a period of 122 days (the number of days in the wet season considered in one particular year). More precisely, for each fixed location $\boldsymbol{s} \in \mathbb{S}$, we deseasonalize the time series $\left\{X(\S,t), t\in \{t_1,\ldots,t_{732}\} \right\}$ by computing for $i=1,\ldots,122$ 
	\begin{equation}
	\tilde{X}(\S,t_{i+122(j-1)}) = {X}(\S,t_{i+122(j-1)})- \frac{1}{6} \sum_{j=1}^{6} {X}(\S,t_{i+122(j-1)}), 
	\end{equation}
	\item For each fixed location $\boldsymbol{s} \in \mathbb{S}$, the deseasonalized observations are fitted to the generalized extreme value distribution, 
	\begin{equation}
	\text{GEV}_{\mu(\boldsymbol{s}),\sigma(\boldsymbol{s}),\xi(\boldsymbol{s})}(x)=\exp \left\{  - \left[1+\xi (\boldsymbol{s}) \left(\frac{x-\mu(\boldsymbol{s})}{\sigma{(\boldsymbol{s})}} \right)  \right]^{-1/\xi (\boldsymbol{s})}\right\}, 
	\end{equation}
	for some location $\mu(\boldsymbol{s}) \in \mathbb{R}$, scale $\sigma(\boldsymbol{s})>0$, and shape $\xi (\boldsymbol{s})\in \mathbb{R}$. Let us remark that the estimated shape parameters $\xi(\S)$ are sufficiently close to zero with confidence interval containing zero, see Figure~\ref{RADARFIT}. This suggests a Gumbel distribution (GEV with $\xi=0$) as appropriate model. Therefore, we fit directly a Gumbel distribution $$\text{GEV}_{\mu(\S),\sigma(\S),0}(x)= \left\{ \exp \left[ - \exp \left( - \frac{x-\mu(\S)}{\sigma(\S)} \right) \right]   \right\}.$$ For each spatial location, we assess the goodness of the marginal fits by QQ-plots of deseasonalized rain series versus the fitted Gumbel distribution. The results at four spatial locations $(1,1), (6,5), (7,10) \ \text{and}\ (11,8)$ are summarized in Figure~\ref{qqplotGumbel}. All plots provide a reasonable fit.
	\item The deseasonalized observations may be transformed either to standard Gumbel or standard Fr\'{e}chet margins. More precisely, let $\widehat{\mu}(\S)$, $\widehat{\sigma}(\S)$ are the parameter estimates obtained from (ii), then we may use:
	\begin{enumerate}
		\item  $\tilde{\tilde{X}}(\S,t) = \frac{{\tilde{X}}(\S,t)-\widehat{\mu}(\S)}{\widehat{\sigma}(\S)}, \ t \in \{1,\ldots,732\}$ to transform the deseasonalized observations to standard Gumbel margins;
		\item $ \tilde{\tilde{X}}(\S,t) = -\frac{1}{\log\left\{ \text{GEV}_{\widehat{\mu}(\S),\widehat{\sigma}(\S),0}({\tilde{X}}(\S,t)) \right\}}, \ t \in \{1,\ldots,732\}$ to transform the deseasonalized observations to standard Fr\'{e}chet margins. This transformation is called the probability integral transformation.	In this study, we adopt this case.
	\end{enumerate}
	
\end{enumerate}

In \cite{buhl2016semiparametric, davis2013statistical}, the authors assume that the observations $\tilde{\tilde{X}}(\S,t)$ are realizations from the space-time max-stable BR process. The contribution of the present section is to broaden the dependence structure by considering the spectrally separable space-time max-stable processes, that allow interactions between spatial and temporal components.    

In the sequel, we estimate the extremal dependence structure for the daily maxima of rainfall measurements. Based on our findings in the simulation studies, we notice that Scheme 1 outperforms Scheme 2 generally. So, one may first estimate the extremal dependence parameters using Scheme 2. Afterward, re-estimating the parameters using Scheme 1, where the estimates resulting from Scheme 2 serve as starting values for the optimization routine used in Scheme 1. To that aim, we consider the following five spatio-temporal max-stable models: 
\begin{enumerate}[label=(\roman*)]
	\item Class A: consists of two non-spectrally separable models A$_1$ and A$_2$.
	\begin{itemize}
		\item A$_1$: a space-time max-stable BR model (\ref{BR spectral fun}), with dependence function $\gamma(\H,l) =  2 \phi_s \HH^{\kappa_s}+2 \phi_t {l'}^{\kappa_t}$, where $l'=\lvert l \rvert$, recall Example~\ref{example1}.

		\item A$_2$: a space-time max-stable Schlather model. The space-time correlation function is chosen to be separable such that
		$$\rho(\H,l)=\exp\left\{- \left[   \left(\HH/\phi_s \right)^{\kappa_s} +\left(l'/\phi_t \right)^{\kappa_t}  \right]\right\},$$
		where the range parameters $\phi_t, \phi_s>0$ and the smoothing parameters $0<\kappa_t, \kappa_s<2$.
	\end{itemize}
	
	\item Class B: consists of spectrally separable models B$_1$, B$_2$ and B$_3$.
	\begin{itemize}
		\item  B$_1$: a spectrally separable space-time max-stable model (\ref{model space-time}), where the innovation process $H$ is derived from independent replications of a spatial BR process with semivariogram $\gamma(\H)= \left(\HH/\phi \right)^\kappa$, for some range parameter $\phi>0$ and smoothness parameter $\kappa \in (0,2]$. Obviously, models A$_1$ and B$_1$ are equivalent when the time lag $l'=0$.
		
		\item B$_2$: a spectrally separable space-time max-stable model (\ref{model space-time}), where the innovation process $H$ is derived from independent replications of a spatial Smith process with covariance matrix $\boldsymbol{\Sigma}=\begin{pmatrix}
		\sigma_{11}&\sigma_{12} \\
		\sigma_{12} &\sigma_{22}
		\end{pmatrix},$ recall Example~\ref{example2}.

		\item B$_3$: a spectrally separable space-time max-stable model (\ref{model space-time}), where the innovation process $H$ is derived from independent replications of a spatial extremal-$t$ process with degrees of freedom $\nu \geq 1$ and correlation function of type powered exponential defined, for all $\HH \geq0$, by $\rho(\H)=\exp [- ({\HH} / \phi)^{\kappa}]$, $\phi>0$ and $0<\kappa<2$, where $\phi$ and $\kappa$ denote, respectively, the range and the smoothing parameters. 
	\end{itemize}

\end{enumerate}	

To select the best-fitting model, we use the Akaike Information Criterion (AIC) which was first developed by \cite{akaike1974new} under the framework of maximum likelihood estimation. The AIC is one of the most widely used methods for selecting a best-fitting model from several competing models given a particular dataset. A concise formulation of the AIC under the framework of least squares estimation has been derived by \cite{banks2017aic}. The AIC under Scheme 1 is defined as 
\begin{equation} \label{AIC}
\text{AIC}_{\text{NLS}}= \lvert\mathcal{H}\rvert \log \left( \frac{\mathcal{L}(\widehat{{\boldsymbol{\psi}}}^{(s)})}{ \lvert\mathcal{H} \rvert}\right)+2(k_s+1) + \lvert\mathcal{K}\rvert \log \left( \frac{\mathcal{L}(\widehat{{\boldsymbol{\psi}}}^{(t)})}{\lvert\mathcal{K} \rvert}\right)+2(k_t+1),
\end{equation}
where $\mathcal{L}(\widehat{{\boldsymbol{\psi}}}^{(s)})$ and $\mathcal{L}(\widehat{{\boldsymbol{\psi}}}^{(t)})$ are the estimated objective functions in space and time with $\omega^{\H}=\omega^{l'}=1$, i.e., 
$$
\mathcal{L}( \widehat{{\boldsymbol{\psi}}}^{(s)} )=  \sum_{\Vert\boldsymbol{h}\Vert =h \in \mathcal{H}}   \left({\widehat{\nu}}_{F} (\boldsymbol{h})- \nu_{F}^{(\S)} (\boldsymbol{h}, \widehat{{\boldsymbol{\psi}}}^{(s)})
\right)^{2}, \ h \in \mathcal{H},	
$$
$$\mathcal{L}( \widehat{{\boldsymbol{\psi}}}^{(t)} )= \sum_{l' \in \mathcal{K}}  \left({\widehat{\nu}}_{F} (l')- \nu_{F}^{(t)} (l',  \widehat{{\boldsymbol{\psi}}}^{(t)} )
\right)^{2}, \ l'\in \mathcal{K}.	
$$
$\lvert\mathcal{\mathcal{A}}\rvert$ denotes the cardinality of the set $\mathcal{A}$, and $k_s$ and $k_t$ are respectively the total number of purely spatial and purely temporal parameters in the underlying model. If $\lvert\mathcal{\mathcal{H}}\rvert / k_s+1 < 40$ and $\lvert\mathcal{\mathcal{K}}\rvert / k_t+1 < 40$, it is suggested to use an adjusted corrected version of $\text{AIC}_{\text{NLS}}$ (\ref{AIC}), see \cite{banks2017aic}, i.e.,  
\begin{equation} \label{AIC1}
\text{AIC}_{{\text{NLS}}_c}= \text{AIC}_{\text{NLS}} + \frac{2(k_s+1) (k_s+2)}{ \lvert\mathcal{\mathcal{H}}\rvert - k_s } + \frac{2(k_t+1) (k_t+2)}{ \lvert\mathcal{\mathcal{K}}\rvert - k_t }.
\end{equation}

Our results are summarized in Table \ref{Fitted model}. Model A$_1$ has the lowest AIC$_\text{NLS}$ value and therefore would be considered as the best candidate for this dataset, closely followed by model B$_3$. Obviously, the temporal estimates ($\widehat{\phi}{_t}$ and $\widehat{\kappa}{_t}$) in the best-fitting model A$_1$ indicate that there is a weak temporal extremal dependence. Recall that the purely temporal $F$-madogram for this model is given by
$$  \nu_{F}^{(t)} (l')=0.5 - \left\{2\Phi\left(\sqrt{  {\phi}{_t}l'^{\kappa_t}}\right)+1\right\}^{-1}, \ l' > 0.$$ 
Accordingly, $ \nu_{F}^{(t)} (l')$ is close to zero for large values of $\phi_t$, indicating asymptotic independence. On the other hand, $ \nu_{F}^{(t)} (l')$ is approximately constant when $\kappa_t$ is small, indicating that the extremal dependence is the same for all $l'$. So, both large $\phi_t$ and small $\kappa_t$ lead to temporal asymptotic independence.

For comparison, we present the semiparametric estimates obtained by \cite{buhl2016semiparametric}; 
$ \widehat{\phi}{_s}=0.3611, \widehat{\kappa}{_s}=0.9876, \widehat{\phi}{_t}=2.3650 \ \text{and} \  \widehat{\kappa}{_t}=0.0818$. On the other hand, the pairwise likelihood estimates obtained by \cite{davis2013statistical} are $ \widehat{\phi}{_s}=0.3485,  \widehat{\kappa}{_s}=0.8858, \widehat{\phi}{_t}=2.4190 \ \text{and} \ \widehat{\kappa}{_t}=0.1973$. Obviously, these estimates are close to our estimates, except the temporal smoothness estimate $\widehat{\kappa}{_t}$ which is relatively large.

\begin{figure} [htp!]
	\centering
	\includegraphics[scale=0.45]{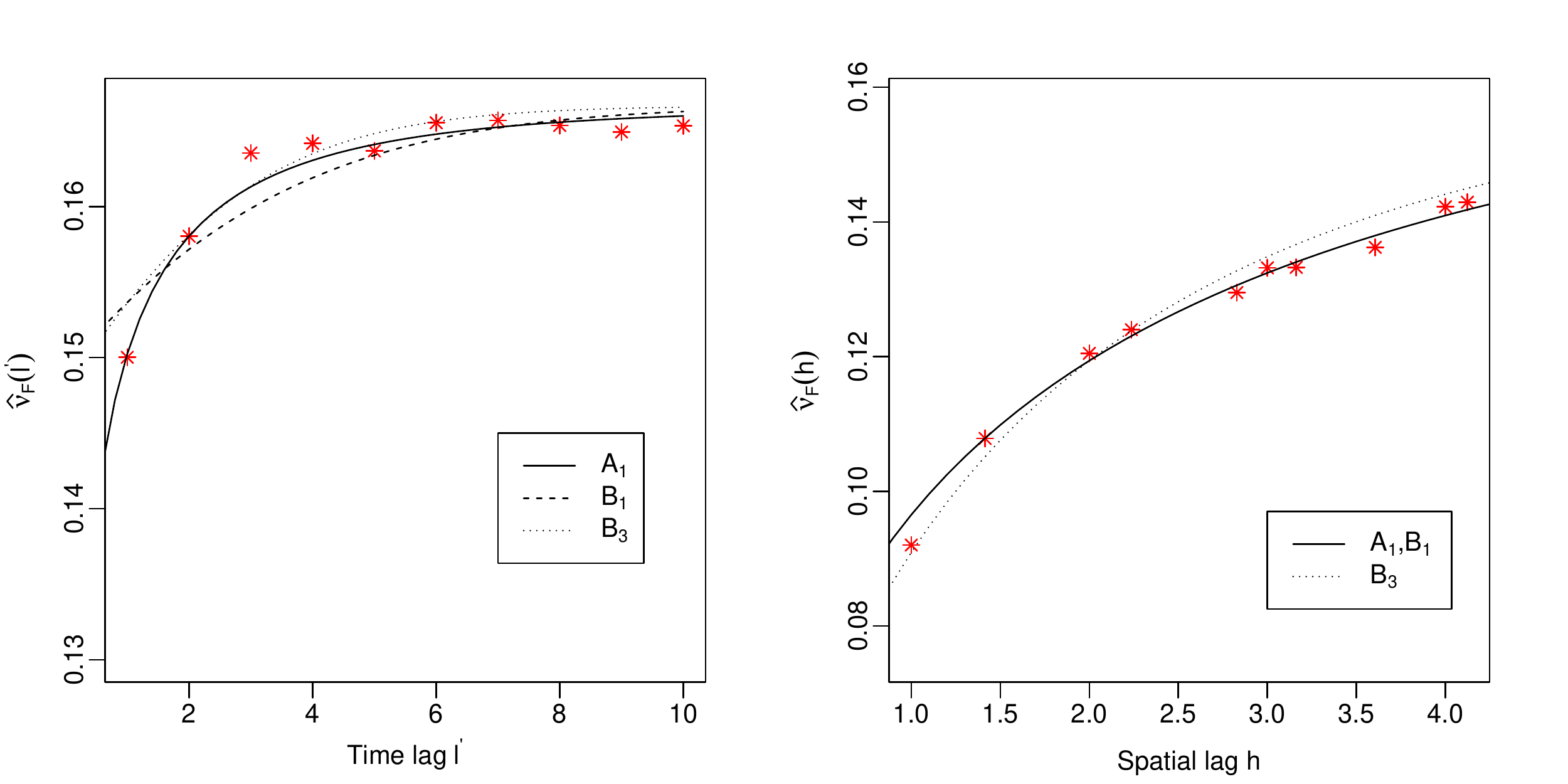}
	
	\caption{Red star symbols show the empirical values of $\nu_{F}(h)$ and $\nu_{F}(l')$ used for estimation. The curves show the fitted $\nu_{F}(h)$ and $\nu_{F}(l')$ from the three best-fitting models (A$_1$, B$_1$ and B$_3$). } 
	
	\label{Goodness_of_fit}
	
\end{figure}

Figure~\ref{Goodness_of_fit} shows the empirical values of $\nu_{F}(h),\ h \in \mathcal{H}$ and $\nu_{F}(l'), \ l' \in \mathcal{K}$, and their model-based counterparts from the three best-fitting models according to the AIC$_\text{NLS}$. It seems that the three models give a quite reasonable fit with a little outperformance for model A$_1$. So, considering these plots and the AIC$_\text{NLS}$ values there is overall evidence in favor of model A$_1$.

\begin{table*}[h!]
	\centering
	\scalebox{0.9}{
		\begin{tabular}{@{}l|lllll@{}}\hline
			
			Model& Purely spatial parameters  && Purely temporal parameters && $\text{AIC}_{{\text{NLS}}_c}$
			\\ \hline

			A$_1$ &   $\widehat{\phi}{_s}=0.4109, \ $  $\widehat{\kappa}{_s}=0.9527$    &&   $\widehat{\phi}{_t}=2.1686, \ $ $\widehat{\kappa}{_t}=0.5410$        &&$\boldsymbol{-64.3921 }$           \\

			A$_2$ &         $\widehat{\phi}{_s}=2.6023, \ $ $\widehat{\kappa}{_s}=1.2600$       &&  $\widehat{\phi}{_t}=2.1902, \ $ $\widehat{\kappa}{_t}=0.3464$         && $-$43.0257       \\
			\hline
			
			B$_1$  &  $\widehat{\phi}=1.2289, \ $ $\widehat{\kappa}=0.9527$          &&       $\widehat{\boldsymbol{\tau}}=(-0.2990,0.1661)^t$,      && $\boldsymbol{-58.7364}$         \\
			&  &&  $\widehat{\delta}=0.5821$ && \\

			B$_2$ &  $\widehat{\sigma}{_{11}}=3.7253, \ $ $\widehat{\sigma}{_{12}}=-0.4181$,         &&  $\widehat{\boldsymbol{\tau}}=(0.5379,-0.1452)^t$, &&      $-$21.4420          \\
			&  $\widehat{\sigma}{_{22}}=4.2100$   &&$\widehat{\delta}=0.1830$    && \\

			B$_3$  &         $\widehat{\phi}=5.9293, \ $$\widehat{\kappa}=1.2491$,             &&               $\widehat{\boldsymbol{\tau}}=(1.4074, 0.8505)^t$, && $\boldsymbol{-59.7906}$          \\
			& $\widehat{\nu}=6.0820$ &&  $\widehat{\delta}=0.5317$ && \\

			\hline
		\end{tabular}
	}
	
	\caption{Summary of the fitted models based on the block maxima design from the radar rainfall measurements in a region in the State of Florida. }
	\label{Fitted model}
\end{table*}


\begin{figure} [htp!]
	\centering
	\includegraphics[scale=0.45]{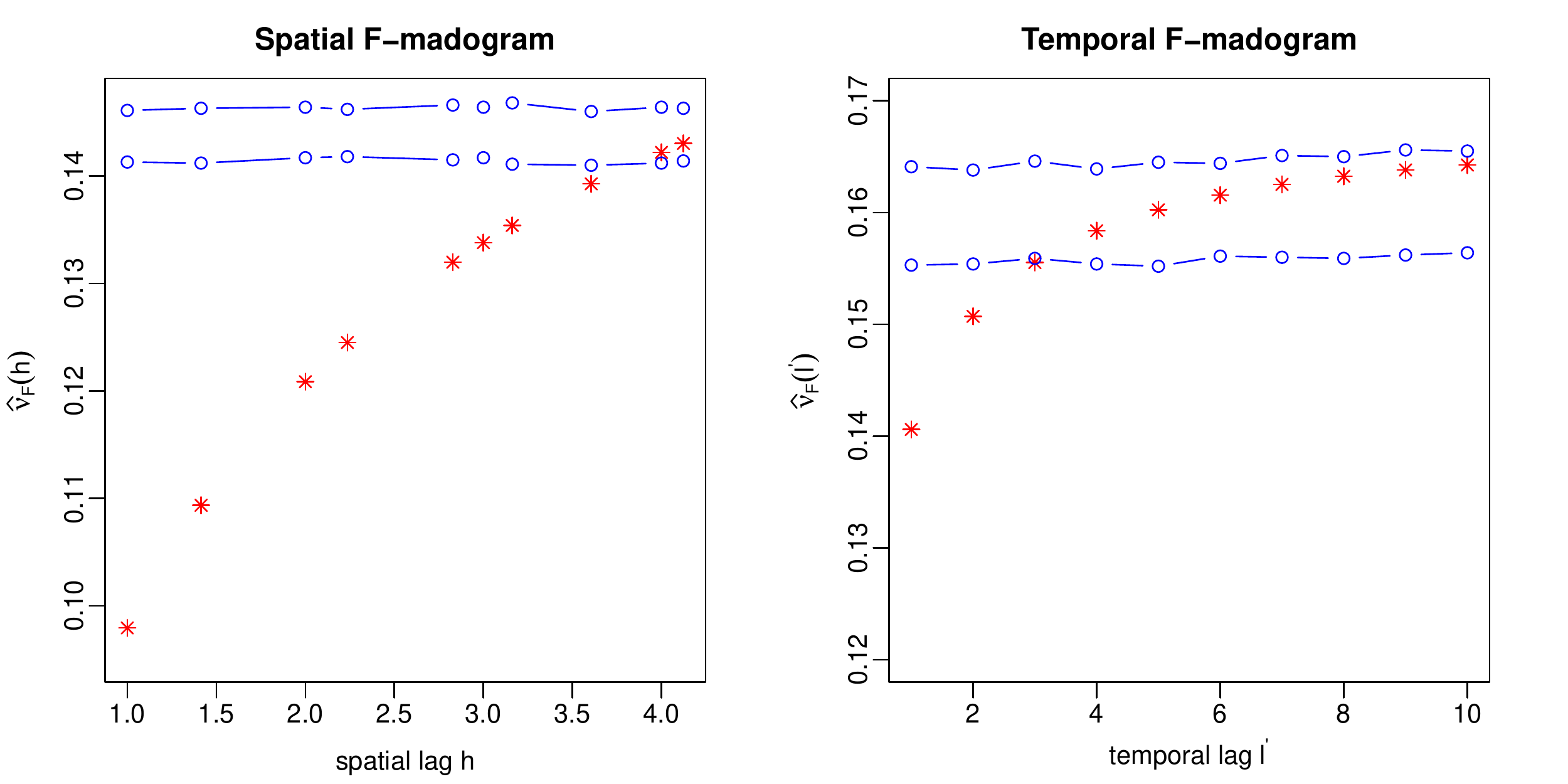}
	
	\caption{Permutation test for extremal independence in space (left panel) and time (right panel). Upper/lower blue lines show 97.5\% and 2.5\% quantiles of empirical $F$-madograms for 1000 spatial (right) and temporal (left) permuations of the space-time observations. Red star symbols show the fitted values of $\nu_{F}(h),\ h \in \mathcal{H}$ and $\nu_{F}(l'), \ l' \in  \mathcal{K}$ derived from the best-fitting model A$_1$. } 
	
	\label{Permutation}
	
\end{figure}

Lastly, permutation tests can be useful to determine the range of clear dependence. So, in order to examine whether the extremal dependence in space and time is significant, we perform a permutation test. We randomly permute the space-time data and compute the empirical spatial/temporal $F$-madograms. More precisely, to check how the extremal dependence lasts in space, for each fixed time point $t \in \{t_1,\ldots, t_{732}\}$ we permute the spatial locations. Afterward, the spatial $F$-madogram is computed and the procedure is repeated 1000 times. From the obtained spatial $F$-madogram sample, we compute 97.5\% and 2.5\% empirical quantiles which form a 95\% confidence region for spatial extremal independence. On the other hand, to test the presence of temporal extremal independence, the analog procedure is done for the temporal $F$-madogram. In particular, for each fixed location $\S \in \mathbb{S} =\left\{(x,y):  x,y \in\{1,\ldots,14\} \right\}$ we sample without replacement from the corresponding time series and compute the empirical temporal $F$-madogram. Our findings are shown in Figure~\ref{Permutation} together with the fitted values of spatial/temporal $F$-madograms derived from the best-fitting models A$_1$. Inspecting these plots, it appears that the spatial extremal dependence vanishes for spatial lags larger than four (the fitted values for the spatial $F$-madogram lies within the obtained independence confidence region), whereas the temporal extremal dependence vanishes for time lags larger than three. Let remark that the same conclusions are obtained in \cite{buhl2016semiparametric}, where the permutation tests have been carried out based on the extremogram.

\section{Concluding remarks} \label{sec:conc}
In summary, motivated by shortcomings in existing inferential methods, we proposed two novel and flexible semiparametric estimation schemes for space-time max-stable processes based on the spatio-temporal $F$-madogram, $\nu_{F}(\H,l)$. Working with the madogram has a few advantages. In addition to its simple definition and the computational facility, it has a clear link with extreme value theory throughout the extremal dependence function. The new estimation procedure may be considered as an alternative or a prerequisite to the widely used pairwise likelihood; the semiparametric estimates could serve as starting values for the optimization routine used to maximize the pairwise log-likelihood function to decrease the computational time and also improve the statistical efficiency, see \cite{castruccio2016high}.

A simulation study has shown that the inference procedure performs well. Moreover, our estimation methodology outperforms the semiparametric estimation procedure suggested by \cite{buhl2016semiparametric} which was based on the dependence measure extremogram. The introduced method is applied to radar rainfall measurements in a region in the State of Florida (Section~\ref{sec:real}) in order to quantify the extremal properties of the space-time observations. 

Our attention is concentrated on fitting space-time max-stable processes based on gridded datasets. In the future, we plan to generalize our method in order to fit space-time max-stable processes with extensions to irregularly spaced locations that may have a fundamental interest in practice. In addition, equally weighted inference approaches have been widely implemented. However, using non-constant weights seems appealing for at least two reasons. First from a computational point of view, for example discarding distant pairs, the CPU load for the evaluation might be smaller and the fitting procedure would be less time-consuming. On the other hand, as neighboring pairs are expected to be strongly dependent, thus providing valuable information for the estimation of dependence parameters, this may improve the statistical efficiency. Therefore, it could be interesting to investigate the gain in statistical efficiency of estimators as well as computational efficiency by adopting different weighting strategies. Since the number of spatial and temporal lags are limited, we could consider weights such that locations and time points which are further apart from each other have less influence on the estimation, i.e.,  
\begin{align*}
\omega^{\H} =& \exp \left\{ -c_1 \HH \right\} \ \text{or}\ \exp \left\{ -c_1 {\HH}^2 \right\} \ \text{or} \ {\HH}^{-c_1} ,\\
\omega^{l'} =& \exp \left\{ -c_2 l' \right\}  \ \text{or}\  \exp \left\{ -c_2 {l'} ^ 2\right\}  \ \text{or}\  l'^{-c_2},\\
\omega^{\H,l'} =& \exp \left\{ -c \left(\HH+l' \right) \right\} \ \text{or}\ \exp \left\{ -c \left({\HH}^2+{l'}^2\right) \right\} \ \text{or} \ {\left(\HH+l' \right)}^{-c} ,
\end{align*}
where $c_1, \ c_2,\  c>0, \ \HH \in \mathcal{H} \ \text{and} \ l' \in \mathcal{K}.$

Finally, it could be interesting to extend the spatial $\lambda$-madogram approach proposed by \cite{naveau2009modelling} to estimate the spatio-temporal extremal dependence function $V_{\H,l}$. For example, in the case of (\ref{model space-time}), it is easy to verify that for $\H \in \R^2$ and $l \in \R$, the spatio-temporal $\lambda$-madogram for any $\lambda \in (0,1)$ is given by
	\begin{equation}
	\nu_{\lambda}(\H,l )=\frac{ (1-\lambda)\{V_{\boldsymbol{0},\boldsymbol{h}-l\boldsymbol{\tau}}\left(\lambda,(1-\lambda) \delta^{-l}\right)\} + {1-\delta^{l}}}{ (1-\lambda)\{1+V_{\boldsymbol{0},\boldsymbol{h}-l\boldsymbol{\tau}}\left(\lambda,(1-\lambda) \delta^{-l}\right)\} + 1-\delta^{l}} - c(\lambda),
	\label{madogram formula1}
	\end{equation}
where $c(\lambda)=\frac{3}{2(1+\lambda)(2-\lambda)}$.

\section*{Acknowledgements}
 We acknowledge the Southwest Florida Water Management District (SWFWMD) for
providing the data. Especially, we would like to thank Margit L. Crowell for the help in finding the data and the related details. We also thank the authors of paper \cite{buhl2016semiparametric} for providing their space-time max-stable BR process simulation {R} code, that used to simulate data in Section \ref{sec:simulation}.

\end{document}